\documentclass[preprint,10pt]{elsarticle}

\usepackage{amsmath, amsthm, mathtools, amssymb, color, float, eqnarray, booktabs, makecell, soul, mdframed, multirow}
\usepackage[hyphens]{url}
\usepackage[justification=centering]{caption}
\usepackage{todonotes}
\usepackage{comment}
\usepackage{bm}
\usepackage{ulem}
\usepackage[bookmarks,colorlinks=true,urlcolor=black,linkcolor=black,citecolor=black]{hyperref}
\usepackage{cleveref}
\usepackage{subcaption}

\usepackage[most]{tcolorbox}

\usepackage[utf8]{inputenc}
\usepackage[english]{babel}
\usepackage{soul}

\journal{arXiv.org}

\begin{document}

\begin{frontmatter}

\title{AI-Aristotle: A Physics-Informed framework for Systems Biology Gray-Box Identification}

\author[label1]{Nazanin Ahmadi Daryakenari\corref{co-first}}
\author[label2]{Mario De Florio\corref{co-first}}
\author[label2]{Khemraj Shukla}
\author[label2]{George Em Karniadakis\corref{corresponding}}

\address[label1]{Center for Biomedical Engineering, School of Engineering, Brown University, Providence, RI, USA}
\address[label2]{Division of Applied Mathematics, Brown University, Providence, RI, USA}

\cortext[co-first]{These authors contributed equally}

\cortext[corresponding]{Corresponding author: george\_karniadakis@brown.edu}

\address{}

\begin{abstract}

Discovering mathematical equations that govern physical and biological systems from observed data is a fundamental challenge in scientific research. We present a new physics-informed framework for parameter estimation and missing physics identification (gray-box) in the field of Systems Biology. The proposed framework -- named AI-Aristotle -- combines eXtreme Theory of Functional Connections (X-TFC) domain-decomposition and Physics-Informed Neural Networks (PINNs) with symbolic regression (SR) techniques for parameter discovery and gray-box identification. We test the accuracy, speed, flexibility and robustness of AI-Aristotle based on two benchmark problems in Systems Biology: a pharmacokinetics drug absorption model, and an ultradian endocrine model for glucose-insulin interactions.
We compare the two machine learning methods (X-TFC and PINNs), and moreover, we employ two different symbolic regression techniques to cross-verify our results. While the current work focuses on the performance of AI-Aristotle based on synthetic data, it can equally handle noisy experimental data and can even be used for black-box identification in just a few minutes on a laptop. More broadly, our work provides insights into the accuracy, cost, scalability, and robustness of integrating neural networks with symbolic regressors, offering a comprehensive guide for researchers tackling gray-box identification challenges in complex dynamical systems in biomedicine and beyond.
\end{abstract}


\end{frontmatter}


\section{Introduction}
One of the most coveted tasks in Machine Learning is the discovery of new physics laws from observed and experimental data. One of the first attempts to extrapolate governing equations from observed data is presented in the well-known work by Brunton et al.  \cite{brunton2016discovering}, in which the authors propose a new school of thought for dynamical system discovery problem from the perspective of sparse regression \cite{tibshirani1996regression} and compressed sensing \cite{donoho2006compressed}. In particular, they take advantage of the fact that most physical systems are described by only a few relevant terms governing the dynamics, making the governing equations sparse in a high-dimensional non-linear function space. This method named SINDy -- Sparse Identification of Nonlinear Dynamics -- depends on the choice of the candidate non-linear functions library and the availability and quality of the data. Thus, it is not a generalized method and works better if guided by available knowledge of the phenomena under study. For example, given the trend of the observed data, one can approximately understand if it is a trigonometric or polynomial trend and build the library accordingly. SINDy has shown its capability in identifying non-linear dynamical systems from data without previous assumptions of the forms of the differential equations governing the phenomena. 

Another method to retrieve governing equations from data has been proposed by Udrescu et al. \cite{udrescu2020ai}. In this paper, the authors make use of the symbolic regression (SR), which aim to find a symbolic expression that accurately represents an unknown function based on a given dataset. They developed a novel recursive multidimensional symbolic regression algorithm, named \textit{AI-Feynman}, that combines neural network techniques with physics-inspired strategies. The efficiency of this method has been proved by discovering 100 equations from the \textit{Feynman Lectures on Physics}, outperforming the accuracy of the state-of-the-art publicly available software. However, despite the groundbreaking capability of this work, there are some drawbacks and areas for improvement. The method currently focuses on equations involving elementary functions but does not handle equations involving derivatives and integrals commonly found in physics. Integrating the capability to discover such equations would be valuable. Also, while the AI-Feynman shows promise, it could further benefit from combining the strengths of genetic algorithms and its approach to generate a more robust and versatile equation discovery tool. Overall, the development and refinement of symbolic regression algorithms continue to evolve, offering exciting possibilities for future discoveries in the realm of physics and beyond. 

On this research direction, a new framework named \textit{AI-Descartes} has been recently published \cite{cornelio2023combining}. In this paper, the authors address the challenge of deriving meaningful mathematical models from both axiomatic knowledge and experimental data by combining logical reasoning with SR. The novelty of this method lies on the attempt  to generate models that are consistent with general logical axioms. The authors showcase the effectiveness of their method by applying it to three classic scientific laws: Kepler's third law of planetary motion, Einstein's relativistic time-dilation law, and Langmuir's theory of adsorption. They demonstrate the capability to discover governing laws even with limited data points, emphasizing the importance of logical reasoning in distinguishing between candidate formulas with similar data-fit accuracy. However, this method relies on the correctness and completeness of background theories, which may not always hold and the development of further techniques such as abductive reasoning \cite{douven2022art} for partially addressing incomplete theories would be needed. Scaling behavior remains a challenge, especially regarding the undecidability of certain logical types and variations in run-time performance. 

Another recently developed SR package, named \textit{Feyn} \cite{brolos2021approach} and based on the symbolic regressor \textit{QLattice}, is showing great performance and capabilities especially for small data sets, where traditional machine learning techniques such as gradient boosting and random forests tend to overfit \cite{wilstrup2021symbolic}. Christensen et al. \cite{christensen2022identifying} efficiently used Feyn on clinical omics datasets to generate high-performing models to predict disease outcomes and to reveal putative disease mechanisms.

Few attempts have been made for discovering ``gray-box'' terms in Differential Equations (DEs) by tackling both data and available known physics. One of these used Physics-Informed Neural Networks (PINNs) \cite{raissi2019physics} with the eXtended-PINN (X-PINN) framework \cite{jagtap2021extended} coupled with SR \cite{kiyani2023framework}. The authors proposed to augment the original X-PINN framework by enforcing a flux continuity across the sub-domains interfaces, and moreover they employed SR to learn the explicit expression of the gray-box unknown function retrieved with the augmented X-PINN. the method proved its effectiveness with the Allen-Cahn equation, whose results indicating excellent performance, even in presence of noise in real-world data scenarios.

Other approaches using particular type of Neural Networks called Random Projection Neural Networks (RPNN) \cite{andras2018random} are used in combination with SR. In a recent work \cite{de2021interaction}, RPNN are used to model a representation for SR called Interaction Transformation \cite{de2018greedy}, showing the capability of this framework in drastically reducing the computational effort. In another work \cite{kokturk2021symbolic}, a single-layer NN is combined with SR. In this approach, the SR layer, incorporating mathematical operators and basis functions, is constructed randomly instead of using genetic programming, and the output weighting parameters are optimized through least-squares optimization. The use of least-squares optimization significantly reduced computational time, resulting in system models based on simple analytic expressions that accurately represent the input-output relationship of dynamic systems. 

One of the earliest works on addressing ``gray-box'' identification for non linear dynamical systems is the one of Ref. \cite{rico1994continuous}. The gray-box in this paper is composed by a known part, represented by a system of Ordinary Differential Equations (ODEs), and unknown parts, which are approximated using neural networks. The paper illustrates this approach by applying it to model a complex reacting system with nonlinear kinetics for parameter discovery. The authors also highlight the challenges of working with discrete-time models and the advantages of using continuous-time approximations for a more nuanced understanding of system behavior. Other gray-box identification and parameter estimation methodologies were applied to a wide range of applications, such as phase field systems, biotechnology and optogenetics \cite{kemeth2023black,lovelett2019partial,quach2007estimating,wandy2018shinykgode,loos2018hierarchical}.

The PINN frameworks are advancing the state-of-the-art of methodologies for inverse problems of parameter discovery. Particularly challenging is the scenario in which we have a highly nonlinear dynamics system with many unknown parameters and very few available experimental data to leverage. This challenge has been addressed in a systems-biology-informed deep learning algorithm that incorporates the system of ODEs into the neural networks. In the works \cite{yazdani2020systems,daneker2023systems}, the authors proved the efficiency of this new algorithm to infer the dynamics of unobserved species using only a few scattered and noisy measurements by testing it for benchmark problems in systems biology.

In this work, we propose a new framework named \textit{AI-Aristotle} to perform parameter discovery and gray-box identification for problems in Systems Biology. We employ two neural networks based methods for the unknown terms approximation, such as PINNs and X-TFC \cite{schiassi2021extreme} with domain decomposition \cite{de2022physics}, and two symbolic regression algorithms for the mathematical explicitation of the gray-box model, such as PySR \cite{virgolin2022symbolic, cranmer2023interpretable} and gplearn \cite{stephens2015gplearn}. Our framework is tested for two problems. The first one is a three-compartment pharmacokinetics model, describing a single-dose drug absorption. The second, more challenging problem, is an ultradian endocrine model, describing the glucose-insulin interaction.\\
This paper is organized as follows. In section \ref{sec:models}, we present an introduction of the physics-based models used for our simulations. In Section \ref{sec:methodology} we report the two Neural Networks methods used for solving the inverse problem with data and physics models, and the two SR algorithms used to explicitly identify the gray-boxes previously retrieved. In In Section \ref{sec:results} we report the results obtained by the two NN methods and the two SR algorithms for different test cases, involving both parameter discovery and gray-box identification. Finally, we summarize conclusions and discussion about in Section \ref{sec:conclusions}.

\section{Models}\label{sec:models}

In this section, the mathematical models describing the phenomena of our simulations are introduced. These models are designed to capture the dynamic interactions within specific biological processes, such as drug absorption and glucose-insulin interaction, offering physics-based knowledge of the behavior and characteristics of the systems under study.


\subsection{Pharmacokinetics model}

The first model we aim to use for our simulations is a single-dose compartmental Pharmacokinetics (PK) model \cite{barnes2011mathematical}, represented by the following system of ODEs:
\begin{equation}\label{eq:PK}
    \begin{cases}
        \dfrac{dB}{dt} = k_g G - k_b B\\
        \dfrac{dG}{dt} = -  k_g G\\
        \dfrac{dU}{dt} =  k_b B
    \end{cases} 
    \qquad \qquad \text{s.t.} \qquad \qquad 
    \begin{cases}
        B(0) = 0 \\
        G(0) = 0.1 \mu g \\
        U(0) = 0
    \end{cases}
\end{equation}
This model evaluates the variation of drug concentration in three compartments, in a time range [0, 10] hours. The drug is initially introduced in the GI-tract (first compartment $G$), where it dissolves and diffuses into the bloodstream (second compartment $B$). Finally, the drug is eliminated from the bloodstream through the liver, kidneys, and urinary tract (third compartment $U$). The parameters $k_g = 0.72 h^{-1}$ and $k_b = 0.15h^{-1}$ represent the rates at which the drug diffuses from the GI-tract into the bloodstream, and then eliminated from the bloodstream through the liver, kidneys, and urinary tract, respectively. The intake drug is considered to be $0.1 \mu g$ of antibiotic tetracycline. In Section \ref{sec:results}, we will show our simulations using this model for two test cases: 1) Parameters discovery, and 2) gray-box identification. With ``gray box'', we indicate the missing terms of a model. For this PK model, the missing term considered is the right-hand-side of the first ODE, which we approximate with an unknown function $h(t)$ as follows:
\begin{equation}\label{eq:pharmacok_f}
    \begin{cases}
        \dfrac{dB}{dt} = h(t)\\
        \dfrac{dG}{dt} = -  k_g G\\
        \dfrac{dU}{dt} =  k_b B
    \end{cases} 
    \qquad \qquad \text{s.t.} \qquad \qquad 
    \begin{cases}
        B(0) = 0 \\
        G(0) = 0.1 \mu g \\
        U(0) = 0
    \end{cases}
\end{equation}
which we aim to obtain by using available data for $B$, $G$, and $U$.


\subsection{Ultradian Endocrine model}

The second model used in our simulations is an ultradian model for the glucose-insulin interaction \cite{sturis1991computer}, which is modeled by 6 state variables and 30 parameters \cite{yazdani2020systems}. This model describes the existence of rhythmic oscillations in both glucose and insulin levels within the body that occur on a relatively short timescale, typically less than 24 hours. In particular, in our simulation we will use a time range [0, 1800] minutes. It results in the following system of ODEs:
\begin{equation}
       \begin{cases}
        \dfrac{dI_p}{dt} = f_1(G) - E \left( \dfrac{I_p}{V_p} - \dfrac{I_i}{V_i} \right) - \dfrac{I_p}{t_p}  \\
        \dfrac{dI_i}{dt} =  E \left( \dfrac{I_p}{V_p} - \dfrac{I_i}{V_i} \right) - \dfrac{I_i}{t_i} \\
        \dfrac{dG}{dt} =  f_4(h_3) + I_G(t) - f_2(G) - f_3(I_i)G \\
        \dfrac{dh_1}{dt} = \dfrac{1}{t_d}(I_p - h_1)  \\
        \dfrac{dh_2}{dt} =  \dfrac{1}{t_d}(h_1 - h_2) \\
        \dfrac{dh_3}{dt} =   \dfrac{1}{t_d}(h_2 - h_3)
    \end{cases}
    \qquad \qquad \text{s.t.} \qquad \qquad 
    \begin{cases}
        I_p(0) = 36 \mu U/ml \\
        I_i(0) = 44 \mu U/ml \\
        G(0) = 110 mg/dl \\
        h_1(0) = 0 \\
        h_2(0) = 0 \\
        h_3(0) = 0         
    \end{cases}
\end{equation}
The three main variables of this model are the plasma insulin concentration $I_p$, the interstitial insulin concentration $I_i$, and the glucose concentration $G$. The last three variables $h_1,h_2,$ and $h_3$ -- a three-stage linear filter -- represent the delay process between insulin and glucose production \cite{sturis1991computer}. The functions $f_1,f_2,f_3,$ and $f_4$, represent the insulin secretion, the insulin-independent glucose utilization, the insulin-dependent glucose utilization, and insulin-dependent glucose utilization, respectively \cite{albers2014dynamical}, and they are expressed as follows:
\begin{align*}
    f_1(G) &= \frac{R_m}{1 + \exp\left(-\frac{G}{V_gC_1} + a_1\right)}, \\
    f_2(G) &= U_b \left(1 - \exp\left(-\frac{G}{C_2V_g}\right)\right), \\
    f_3(I_i) &= \frac{1}{C_3V_g} \left(U_0 + \frac{U_m}{1 + (\kappa I_i)^{-\beta}}\right), \\
    f_4(h_3) &= \frac{R_g}{1 + \exp\left(\alpha \left(\frac{h_3}{C_5V_p} - 1\right)\right)},
\end{align*}
where
\begin{equation*}
    \kappa = \frac{1}{C_4} \left(\frac{1}{V_i} + \frac{1}{Et_i}\right),
\end{equation*}
and $I_G(t)$ is the exogenous (externally driven) glucose delivery rate. In our simulations, we define it over $N=3$ nutrition events, at time $t_j$ with a carbohydrate quantity $m_j$:
\begin{equation}
    I_G(t) = \sum_{j=1}^{N} m_jk \exp(k(t_j - t)), 
\end{equation}
where $(t_j,m_j) = [(300, 60) (650, 40) (1100)] (min,g)$.
The parameters governing this system of ODEs are listed in Table \ref{tab:param}.
\begin{table}[ht]
\centering
\begin{tabular}{cccc}
\hline
\textbf{Parameter} & \textbf{Nominal value} & \textbf{Unit} & \textbf{Search range} \\
\hline
\hline
$V_p$ & 3 & $lit$ &   --\\
$V_i$ & 11 & $lit$ &  -- \\
$V_g$ & 10 & $lit$ &  -- \\
$E$ & 0.2 & $lit \, min^{-1}$ & $(0.1, 0.3)$ \\
$t_p$ & 6 & $min$ & (4, 8) \\
$t_i$ & 100 & $min$ & (60, 140) \\
$t_d$ & 12 & $min$ &  -- \\
$k$ & 0.0083 & $min^{-1}$ &  -- \\
$R_m$ & 209 & $mU \, min^{-1}$ & $(41.8, 376.2)$ \\
$a_1$ & 6.6 & --  &  $(1.32 , 11.88)$\\
$C_1$ & 300 & $mg \, lit^{-1}$ &  -- \\
$C_2$ & 144 & $mg \,lit^{-1}$ &  -- \\
$C_3$ & 100 & $mg \,lit^{-1}$ &  -- \\
$C_4$ & 80 & $mU \,lit^{-1}$ &  -- \\
$C_5$ & 26 & $mU \,lit^{-1}$ &  -- \\
$U_b$ & 72 & $mg \,min^{-1}$ &  -- \\
$U_0$ & 4 & $mg \,min^{-1}$ &  -- \\
$U_m$ & 90 & $mg \,min^{-1}$ &  -- \\
$R_g$ & 180 & $mg\, min^{-1}$&  --  \\
$\alpha$ & 7.5 & -- &  -- \\
$\beta$ & 1.772 & -- &  -- \\
\hline
\hline
\end{tabular}
\caption{ Ultradian Endocrine model: List of parameters for the model. The search ranges are listed only for the five parameters used for the parameter discovery in our simulations.}\label{tab:param}
\end{table}
Also for this second model, we aim to pursue parameter discovery and gray-box identification. For the latter case, the missing terms we approximate with two unknown functions, $f(t)$ and $g(t)$, whih are in the first two ODEs, as follows:
\begin{equation}\label{eq:GI_system_f_t}
    \begin{cases}
        \dfrac{dI_p}{dt} = f_1(G) + f(t)  \\
        \dfrac{dI_i}{dt} =  g(t) \\
        \dfrac{dG}{dt} =  f_4(h_3) + I_G(t) - f_2(G) - f_3(I_i)G \\
        \dfrac{dh_1}{dt} = \dfrac{1}{t_d}(I_p - h_1)  \\
        \dfrac{dh_2}{dt} =  \dfrac{1}{t_d}(h_1 - h_2) \\
        \dfrac{dh_3}{dt} =   \dfrac{1}{t_d}(h_2 - h_3).
    \end{cases}
\end{equation}

\section{Methodology}\label{sec:methodology}

As mentioned in the Introduction section, the parameter discovery and approximation of the unknown terms in the systems of ODEs are performed by two NN-based methods, while the symbolic regression is performed by two different algorithms, to cross-verify the mathematical expressions obtained. In this section, we present some details of these methods that are included in the AI-Aristotle framework. 

\subsection{X-TFC}

The first NN-based method presented uses a single-layer random projection neural network. For the sake of simplicity, we will show its implementation for the gray-box identification in the pharmacokinetics model only, since the implementation for the ultradian endocrine model is similar. 

\begin{figure}[h!]
    \centering
    \includegraphics[width=\linewidth]{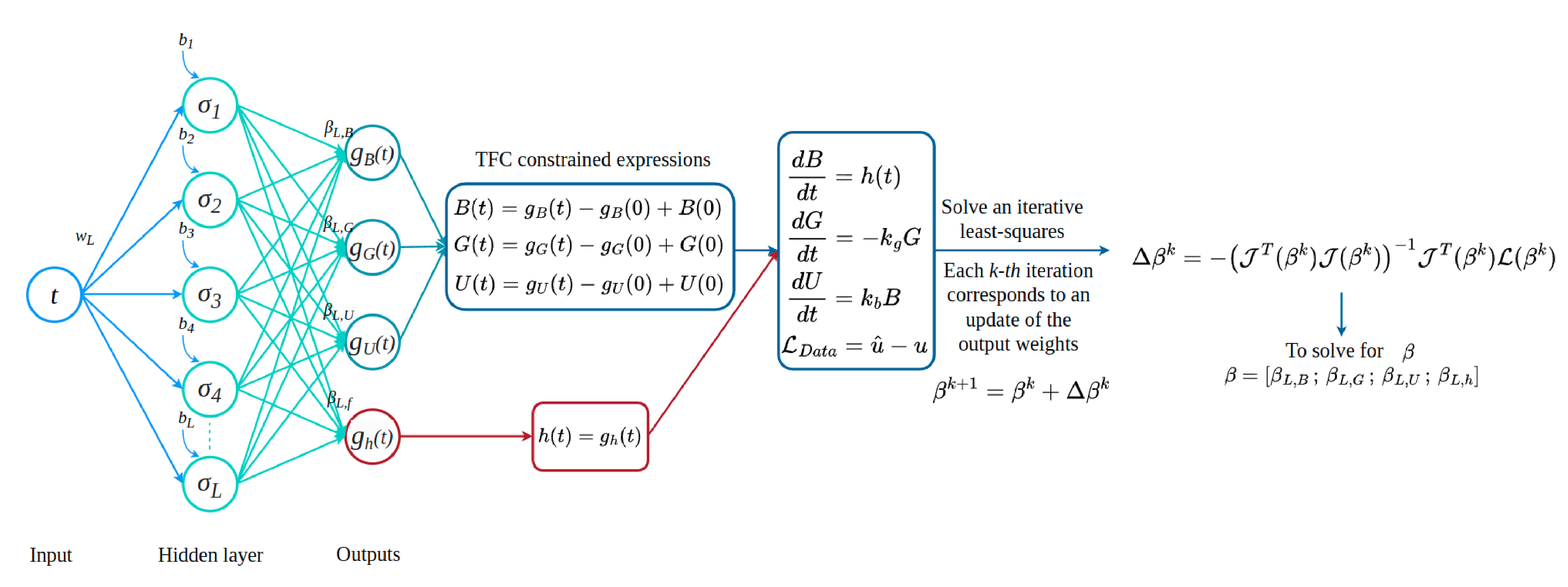}
    \caption{Pharmacokinetics model: Schematic of the X-TFC algorithm. Input weights and biases are randomly selected. The last step solves iteratively a least squares system, thus no back-propagation is involved in the training, allowing fast computational times.}
    \label{fig:schematic_xtfc}
\end{figure}

Different techniques are combined to build this algorithm for solving both forward and inverse problems involving differential equations. The first one is a functional interpolation technique named Theory of Functional Connections (TFC) \cite{mortari2017theory,de2021theory}. According to TFC \cite{mortari2017least}, we can approximate the unknown solutions of our system of ODEs, taking into consideration the initial conditions, with the so-called constrained expressions (CE) as follows:
\begin{subequations}
\label{eq:CEs}
\begin{align}
    B &= \bigl( \boldsymbol{\sigma}(t) - \boldsymbol{\sigma}(0) \bigr) ^T \boldsymbol{\beta}_{B} + B(0) \label{eq:CEs1}\\
    G &= \bigl( \boldsymbol{\sigma}(t) - \boldsymbol{\sigma}(0) \bigr) ^T \boldsymbol{\beta}_{G} + G(0) \label{eq:CEs2}\\
    U &= \bigl( \boldsymbol{\sigma}(t) - \boldsymbol{\sigma}(0) \bigr)^T  \boldsymbol{\beta}_{U} + U(0) \label{eq:CEs3}
\end{align}
\end{subequations}
whose derivatives can be analytically expressed:
\begin{subequations}
\begin{align}
    \dfrac{dB}{dt} &= c\boldsymbol{{\sigma'}} ^T \boldsymbol{\beta}_{B} \\
    \dfrac{dG}{dt} &= c\boldsymbol{{\sigma'}} ^T \boldsymbol{\beta}_{G} \\
    \dfrac{dU}{dt} &= c\boldsymbol{{\sigma'}} ^T \boldsymbol{\beta}_{U}
\end{align}
\end{subequations}
The parameter $c$ represents a mapping coefficient that maps the time domain $t$ into the activation function domain. 
To these systems, we need to add the NN approximation of the unknown term $h(t)$, which is
\begin{equation}\label{eq:CEsf}
    h(t) = \boldsymbol{\sigma}(t)\boldsymbol{\beta}_{h}.
\end{equation}
Here, $\boldsymbol{\sigma}$ is the free-chosen function of the CE. No matter what free-chosen function will be selected, the CE will always satisfy the initial conditions exactly. According to the X-TFC framework \cite{schiassi2021extreme}, we select a single-layer NN as free-chosen function, such as
\begin{equation}\label{eq:singleNN}
g(t) = \sum_{j=1}^{L} \beta_j\sigma \left(w_jt + b_j \right)= \begin{bmatrix}
\sigma(w_1 t + b_1) \\ \vdots \\  \sigma(w_L t + b_L) 
\end{bmatrix}^T  \boldsymbol{\beta} = \boldsymbol{\sigma}^T \boldsymbol{\beta}
\end{equation}
where $L$ is the number of neurons, $w_j \in \mathbb{R}$ is the $j$\textsuperscript{th} input weight connecting the input node with the $j$\textsuperscript{th} neuron, $\beta_j \in \mathbb{R}$ with $j=1,...,L$ is the $j$\textsuperscript{th} output weight connecting the output node with the $j$\textsuperscript{th} neuron, $b_j$ is the bias of the $j$\textsuperscript{th} neuron, and $\sigma_j(\cdot)$ is the NN's activation function, which is selected by the user (for all the simulations in this work, we select a $tanh$ activation function). In the extreme learning machine algorithm \cite{huang2006extreme}, input weights and biases are randomly pre-selected (uniform random distribution), thus the only unknown parameters that need to be computed are the output weights $\boldsymbol{\beta} = \left[ \beta_{1},...,\beta_{L} \right]^T$. 
Once the CEs are built, they can be replaced in the system of ODEs of Eq. \eqref{eq:pharmacok_f}, to obtain the loss functions
\begin{subequations}
\begin{align}
    \mathcal{L}_B &= - c\boldsymbol{{\sigma'}}(t) ^T \boldsymbol{\beta}_{U} + \boldsymbol{\sigma}(t)\boldsymbol{\beta}_{h}  \\
    \mathcal{L}_G &= \Bigl( - c\boldsymbol{{\sigma'}}(t) ^T  - k_g \bigl( \boldsymbol{\sigma}(t) - \boldsymbol{\sigma}(0) \bigr) \Bigr) ^T \boldsymbol{\beta}_{G} + G(0)   \\
    \mathcal{L}_U &=  - c\boldsymbol{{\sigma'}}(t) ^T \boldsymbol{\beta}_{U} + k_b   \bigl( \boldsymbol{\sigma}(t) - \boldsymbol{\sigma}(0) \bigr) ^T \boldsymbol{\beta}_{B} + k_b B(0)  \\
    \mathcal{L}_{data_B} &= \tilde{B} - \bigl( \boldsymbol{\sigma}(t) - \boldsymbol{\sigma}(0) \bigr) ^T \boldsymbol{\beta}_{B} + B(0)    \\
    \mathcal{L}_{data_G} &= \tilde{G} - \bigl( \boldsymbol{\sigma}(t) - \boldsymbol{\sigma}(0) \bigr) ^T \boldsymbol{\beta}_{G} + G(0)  \\
    \mathcal{L}_{data_U} &=  \tilde{U} - \bigl( \boldsymbol{\sigma}(t) - \boldsymbol{\sigma}(0) \bigr) ^T \boldsymbol{\beta}_{U} + U(0), 
\end{align}
\end{subequations}
where $\tilde{B},\tilde{G}$, and $\tilde{U}$ are the available observed data of the three variables. As we can see, now we have reduced the problem into a system of linear equations of the type $Ax=b$, where the unknown $x$ is the vector of output weights $\boldsymbol{\beta}$. However, here we show the procedure to solve it as a system of non-linear equations (that will be the case of the Ultradian Endocrine model). 
When dealing with a system of non-linear ODEs, the next step is to build the Jacobian matrix, by deriving the six previous losses with respect to $\boldsymbol{\beta}_B, \boldsymbol{\beta}_G, \boldsymbol{\beta}_U$, and $\boldsymbol{\beta_f}$. For the pharmacokinetics model, the Jacobian is
\begin{equation}
    \boldsymbol{\mathcal{J}} = \begin{bmatrix}
        \dfrac{\partial \mathcal{L}_B}{\boldsymbol{\beta}_B} & \dfrac{\partial \mathcal{L}_B}{\boldsymbol{\beta}_G} & \textbf{0} & \dfrac{\partial \mathcal{L}_B}{\boldsymbol{\beta}_h}\\
        \textbf{0} & \dfrac{\partial \mathcal{L}_G}{\boldsymbol{\beta}_G} & \textbf{0} & \textbf{0} \\
        \dfrac{\partial \mathcal{L}_U}{\boldsymbol{\beta}_B} & \textbf{0} & \dfrac{\partial \mathcal{L}_U}{\boldsymbol{\beta}_U} & \textbf{0} \\
        \dfrac{\partial \mathcal{L}_{data_B}}{\boldsymbol{\beta}_B} & \textbf{0} & \textbf{0} & \textbf{0}  \\
        \textbf{0} & \dfrac{\partial \mathcal{L}_{data_G}}{\boldsymbol{\beta}_G} & \textbf{0} & \textbf{0}  \\
        \textbf{0} & \textbf{0} & \dfrac{\partial \mathcal{L}_{data_U}}{\boldsymbol{\beta}_U} & \textbf{0}. 
    \end{bmatrix} 
\end{equation}
The unknown vector $\boldsymbol{\beta}$ is computed by iteratively solve the linear system $\boldsymbol{\mathcal{J}} \Delta \beta^k = \boldsymbol{\mathcal{L}}$. Each \textit{k-th} iteration corresponds to an update of the output weights $\beta^{k+1} = \beta^k + \Delta \beta^k$, where $\Delta \beta^k = -\left( \mathcal J^T(\beta^k)\mathcal J(\beta^k)  \right)^{-1} \mathcal J^T(\beta^k)\mathcal L(\beta^k) $.
Once all the output weights $\boldsymbol{\beta}$ are computed, they will be replaced into the CEs of \cref{eq:CEs1,eq:CEs2,eq:CEs3} and eq. \eqref{eq:CEsf} to find our sought solutions. 
In this work, X-TFC is used in a domain-decomposition fashion \cite{de2022physics,schiassi2022physics}, where the time-domain is decomposed into several sub-domains with equispaced time steps, and the algorithm is applied to each sub-domain subsequently, such that the solution found at the interface becomes the new initial condition for the subsequent iteration of the algorithm in the next sub-domain. A schematic of the X-TFC algorithm to solve the gray-box inverse problem for the pharmacokinetics model is shown in Figure \ref{fig:schematic_xtfc}.


\subsection{Physics-Informed Neural Networks (PINNs)}
The second NN-based approach is known as Physics-Informed Neural Networks (PINNs). This method has the capability to address both forward and inverse problems associated with differential equations, by using a deep fully connected neural network.

\subsubsection{PINNs for Pharmacokinetics model}
Building upon the concept of PINNs as originally proposed in reference \cite{raissi2019physics}, we introduce a deep learning framework that incorporates the differential equations governing single-dose compartmental Pharmacokinetics model. In this framework, a neural network characterized by parameters $\theta_{1}$ takes time $t$ as input and generates an output vector representing the state variables $\hat{u}(t; \theta_{1})=(\hat{u}_{B}(t, \theta_{1}),\hat{u}_{G}(t; \theta_{1}), \hat{u}_{U}(t; \theta_{1}))$ which serves as an approximation of the ODE solution $u(t)$.
To solve the gray-box inverse problem, in addition to the unknown parameters, we have an unknown component of the equation. Thus, we introduce another neural network with a different design to approximate the unknown term $h(t)$. The system of ODEs for this model is as follows:
\begin{equation}\label{eq: PINN_graybox}
    \begin{cases}
        \dfrac{dB}{dt} = h(t; \theta_{2})\\
        \dfrac{dG}{dt} = -k_{g}G\\
        \dfrac{dU}{dt} = k_{b}B
    \end{cases} 
    \qquad \qquad \text{s.t.} \qquad \qquad 
    \begin{cases}
        B(0) = 0 \\
        G(0) = 0.1 \mu g \\
        U(0) = 0.
    \end{cases}
\end{equation}

\begin{figure}[h!]
    \centering
    \includegraphics[width=\linewidth]{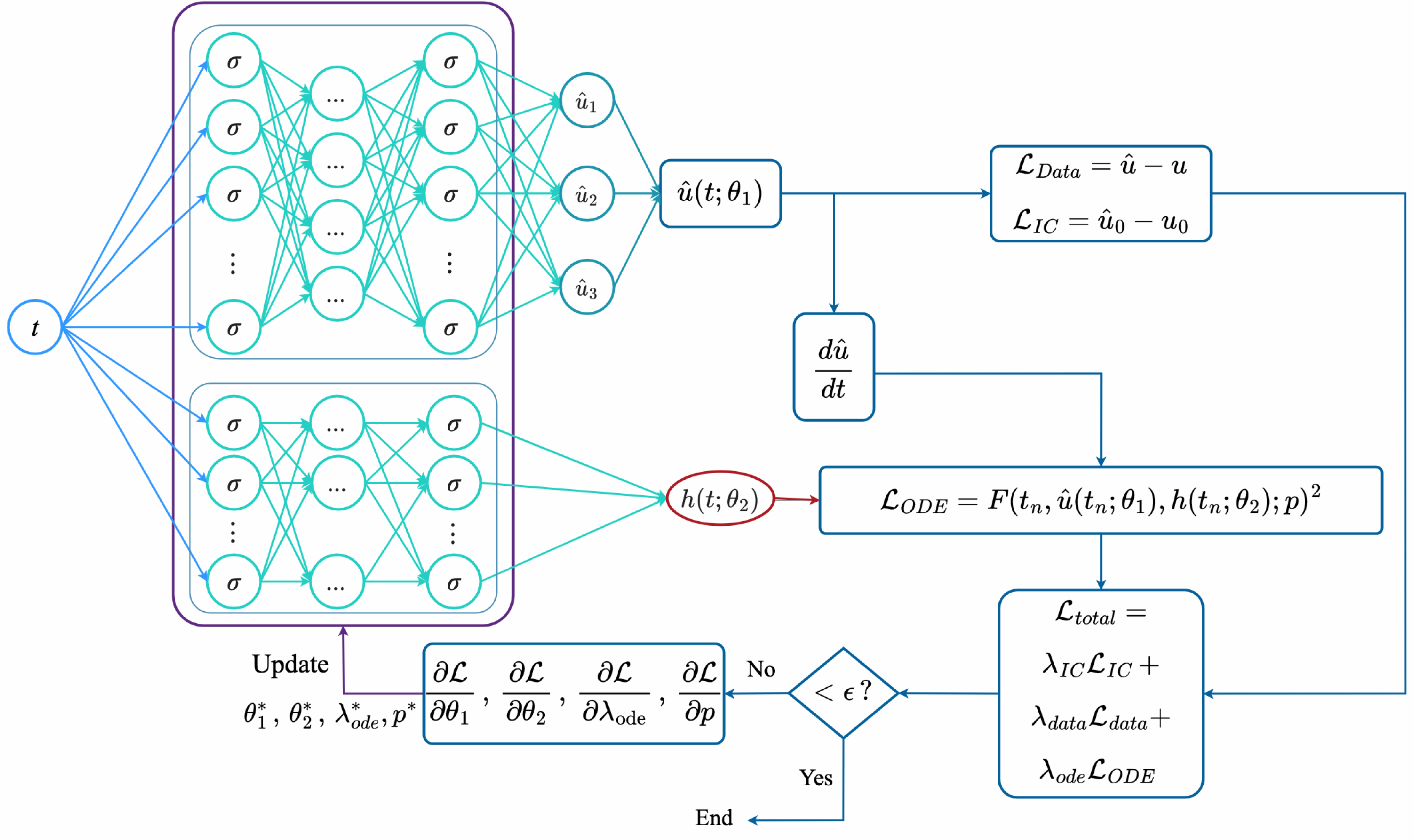}
    \caption{Pharmacokinetics model: Schematic of the PINNs algorithm for predicting the unknown term $h(t;\theta_2)$ and the values of parameters simultaneously. Here, $\hat{u}(t;\theta_1)$ is a vector that contains all three output states. Unlike the X-TFC network, PINNs requires back-propagation, which is the expensive computational component.}
    \label{fig:schmatic_pinn_drug}
\end{figure}

Here, the parameters $\theta_{2}$ characterize the second neural network, which takes $t$ as input and generates an output $h(t; \theta_{2})$.\\
The next crucial step involves constraining the neural network to satisfy both the scattered observations of $u(t)$ and the system of ODEs \eqref{eq: PINN_graybox}. This is achieved by constructing the loss function that takes into account terms corresponding to the observations and the ODE system. To be more specific, let us assume that we have measurements of $u_{\text{data}} = \{u_1, u_2, \ldots, u_M\}$ at various time instances $t_1, t_2, \ldots, t_{M^{\text{data}}}$. We want to ensure that the neural network satisfies the ODE system at specific time points $t_1, t_2, \ldots, t_{N^{\text{ode}}}$. It is important to note that the time instants $t_1, t_2, \ldots, t_{M^{\text{data}}}$, and $t_1, t_2, \ldots, t_{N^{\text{ode}}}$ may not necessarily be on a uniform grid and can be chosen arbitrarily. Here, $N$ is the number of collocation points, and $M$ is the number of data points.\\ For computing the total loss, we employ the Self-Adaptive Loss Balanced method \cite{xzp2022,mcclenny2023selfadaptive}. The total loss function is defined as a function of $\theta_{1}, \theta_{2}, p, \lambda_{\text{ode}}$, where $p$ represents the unknown parameters of the ODEs, and $\lambda_{\text{ode}}$ is a vector representing the individual loss weights for all the state variables, i.e., $\lambda_{\text{ode}} = (\lambda_{1}, \lambda_{2}, \ldots, \lambda_{S})$, where $S$ is the number of state variables. Note that $\lambda_{\text{data}}$ and $\lambda_{\text{IC}}$ are constant values, equal to 1 in this study, and are not trainable variables in our neural network. The total loss is computed as follows:

\begin{equation}
    \mathcal{L}(\theta_{1}, \theta_{2}, p, \lambda_\text{ode}) = \lambda_\text{IC}\mathcal{L_{\text{IC}}}(\theta_{1}) + \lambda_\text{data}\mathcal{L_{\text{data}}}(\theta_{1}) + \lambda_\text{ode}\mathcal{L_{\text{ode}}}(\theta_{1}, \theta_{2}, p), 
\end{equation}

where
\begin{equation}
    \mathcal{L_{\text{IC}}}(\theta) = \left(\hat{u}(t_0) - \hat{u}(t_0; \theta_{1})\right)^2
\end{equation}

\begin{equation}
    \mathcal{L_{\text{data}}}(\theta_{1}) = \frac{1}{M^{\text{data}}} \sum_{m=1}^{M^{\text{data}}} (u(t_m) - \hat{u}(t_m; \theta_{1}))^2
\end{equation}

\begin{equation}
    \mathcal{L_{\text{ode}}}(\theta_{1}, \theta_{2}, p) = \frac{1}{N^{\text{ode}}}\sum_{n=1}^{N^{\text{ode}}} \left(\frac{d\hat{u}}{dt}\bigg|_{t_n} - F(t_n,\hat{u}(t_n; \theta_{1}), h(t_n; \theta_{2}) ; p)\right)^2.
\end{equation}\\
We emphasize that $\mathcal{L_{\text{data}}}$ and $\mathcal{L_\text{IC}}$ represent the discrepancies between the neural network predictions and the measured data, making them supervised losses. Conversely, $L_{\text{ode}}$ is derived from the ODE system and, therefore, qualifies as an unsupervised loss.
In the final step, we simultaneously determine the parameters $\theta^*_{1}$, $\theta^*_{2}$ of both neural networks and the unknown ODE parameters $p^*$ by minimizing the loss function using gradient-based optimization methods, such as the Adam optimizer \cite{kingma2014adam} and L-BFGS optimizer \cite{liu1989limited} . Additionally, we determine the $\lambda^*_\text{ode}$ vector by updating adaptive weights in each epoch by solving:

\begin{equation}
    \theta^*_{1}, \theta^*_{2}, p^*, \lambda^*_\text{ode} = \arg \max_{\lambda_\text{ode}} \min_{\theta_{1}, \theta_{2}, p} \mathcal{L}(\theta_{1}, \theta_{2}, p, \lambda_\text{ode})
\end{equation}\\
For the training process, where our goal is to predict the unknown term $h(t;\theta_2)$ and the values of parameters simultaneously, we employ the Adam optimizer with default hyperparameters and a learning rate of $10^{-4}$. Training is performed on the entire dataset. Since our total loss comprises two supervised losses and one unsupervised loss, we adopt a two-stage training strategy  as follows:

\begin{enumerate}
  \item Recognizing that supervised training typically yields faster convergence than unsupervised training, we initially train the network using the two supervised losses, $\mathcal{L}_{\text{data}}$ and $\mathcal{L}_{\text{IC}}$, for a set number of iterations. This initial training phase enables the network to quickly align with the observed data points.
  \item Subsequently, we continue the training process, incorporating all three losses.
\end{enumerate}
Empirical observations demonstrate that this two-stage training approach expedites network convergence. The specific number of iterations for each stage and parameters for the implementation are detailed in Table \ref{tab:PINN_drug_hyperparam}.
A schematic of the PINNs algorithm for solving the gray-box inverse problem in the pharmacokinetics model is shown in Figure \ref{fig:schmatic_pinn_drug}.

\subsubsection{PINNs for Ultradian Endocrine model}
The system of ODEs for this model is as follows:
\begin{equation}
    \begin{cases}
        \dfrac{dI_p}{dt} = f_1(G) + f(t,\theta_2)  \\
        \dfrac{dI_i}{dt} =  g(t,\theta_2) \\
        \dfrac{dG}{dt} =  f_4(h_3) + I_G(t) - f_2(G) - f_3(I_i)G \\
        \dfrac{dh_1}{dt} = \dfrac{1}{t_d}(I_p - h_1)  \\
        \dfrac{dh_2}{dt} =  \dfrac{1}{t_d}(h_1 - h_2) \\
        \dfrac{dh_3}{dt} =   \dfrac{1}{t_d}(h_2 - h_3)
    \end{cases}
\end{equation}
Here, parameters $\theta_{2}$ characterize the second neural network, which takes $t$ as input and generates two outputs $f(t; \theta_{2})$ and $g(t; \theta_{2})$.\\
In accordance with the pharmacokinetics model, this study adopts a self-adaptive loss balanced method and a two-stage training strategy. To expedite the training process of the neural network, extending the discussion from the previous section on Fully connected Neural Networks, we introduce supplementary layers following the workflow presented in \cite{yazdani2020systems}.

\begin{figure}[h!]
    \centering
    \includegraphics[width=\linewidth]{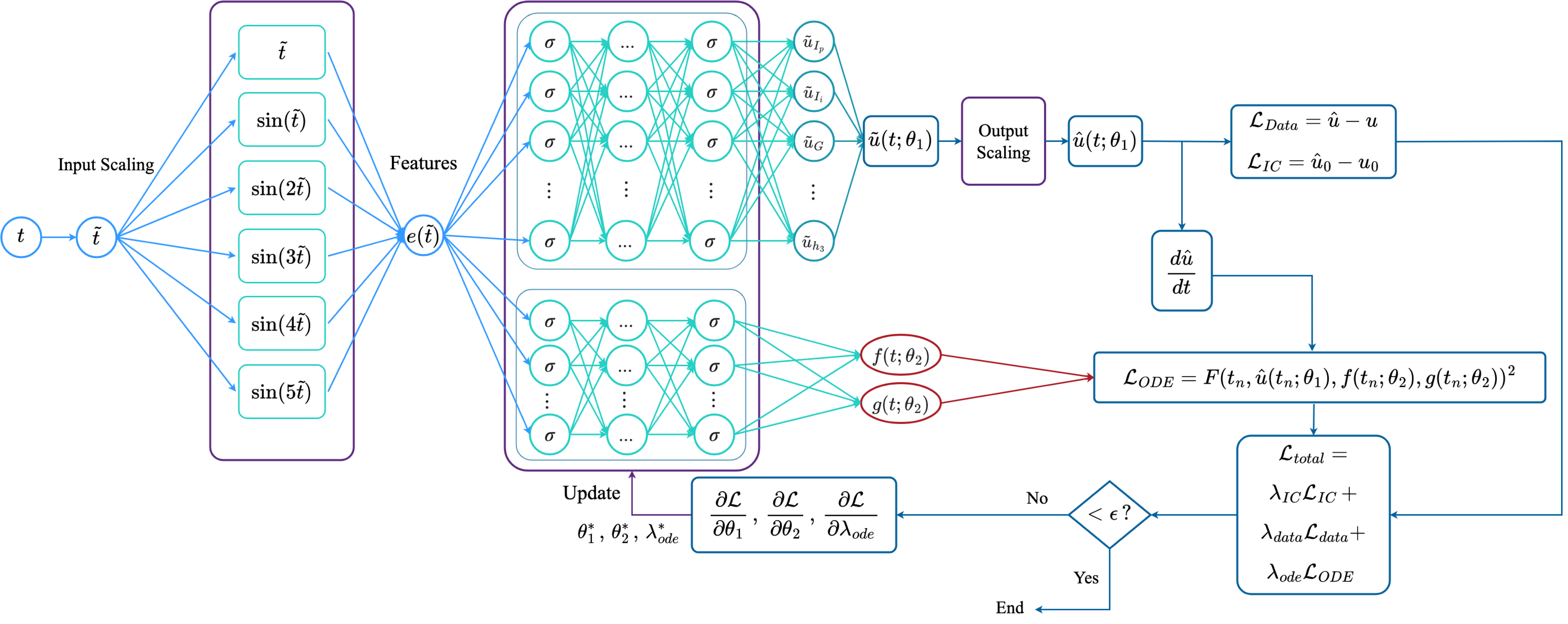}
    \caption{Ultradian Endocrine model: Schematic of the PINNs algorithm for solving a gray-box identification problem.}
    \label{fig:schmatic_pinn_gluc}
\end{figure}

\begin{itemize}
  \item Input Scaling Layer: In cases where the time domain exhibits significant variation spanning multiple orders of magnitude, which can detrimentally affect neural network training, we employ a linear scaling function on the time variable $t$, using a value in the time domain $T$ to obtain $\tilde{t} = \frac{t}{T}$, which approximates values to be $\sim O(1)$. In this study, for the time interval ranging from 0 to 1800, we have adopted a value of $T=100$.
  \item Feature Layer: Frequently, solutions to ordinary differential equations (ODEs) exhibit specific patterns, such as periodicity or exponential decay. Instead of relying on the neural network to autonomously identify these features, we incorporate them within a dedicated feature layer. While the choice of features is problem-specific, the general framework remains consistent across different problems. We utilize the set of functions ${e_1(\theta), e_2(\theta), \ldots, e_L(\theta)}$ to construct $L$ features ${e_1(\tilde{t}), e_2(\tilde{t}), \ldots, e_L(\tilde{t})}$, as illustrated in Figure \ref{fig:schmatic_pinn_gluc}.
  If discerning a clear pattern proves challenging, it is advisable to omit the feature layer rather than introducing inaccurate information. This feature layer is a training aid and not a mandatory component for the success of the PINNs for system biology identification problems.
  \item Output Scaling Layer: The predicted outputs, denoted as $\tilde{u}_{I_p}, \tilde{u}_{I_i}, \ldots, \tilde{u}_{h_3}$, may exhibit variations in magnitudes. To address this, we can normalize the network outputs. To standardize these outputs, we employ a normalization procedure, expressed as follows:
\begin{align*}
\hat{u}_{I_p} &= k_{I_p} \tilde{u}_{I_p} \\
\hat{u}_{I_i} &= k_{I_i} \tilde{u}_{I_i} \\
&\vdots \\
\hat{u}_{h_3} &= k_{h_3} \tilde{u}_{h_3}. 
\end{align*}
Here, $k_{I_p}, k_{I_i}, \ldots, k_{h_3}$ represent the magnitudes of the corresponding ODE solutions $u_{I_p}, u_{I_i}, \ldots, u_{h_3}$. This normalization ensures that the predicted outputs are scaled consistently with the characteristics of the underlying ODE solutions. Furthermore, we introduce an additional component to this layer to facilitate alignment of the state variables with a linear trajectory connecting the initial and final data points. This linear transformation facilitates the interpretation and visualization of the model's outputs, ensuring their alignment with meaningful data trends. In summary, the Output Scaling Layer standardizes predicted outputs while integrating a linear transformation component. This integration enhances the interpretability and relevance of the model's results, expediting the neural network's convergence towards an accurate solution.

\end{itemize}

 The list of parameters of this model can be found in Table \ref{tab:PINN_GI_hyperparam}. A schematic of the PINNs algorithm for solving the gray-box identification problem in the Ultradian Endocrine model is shown in Figure \ref{fig:schmatic_pinn_gluc}.

\subsection{Symbolic Regression}
Symbolic regression is a powerful method used in machine learning, designed to discover a mathematical expression or equation that provides the optimal fit for a provided dataset. Unlike traditional regression methods (e.g., linear regression, polynomial regression), symbolic regression seeks to discover the underlying mathematical relationship between input variables and the target variable without making assumptions about the form of the equation. Two popular symbolic regression algorithms commonly used in this context are PySR (Python Symbolic Regression) \cite{virgolin2022symbolic} and gplearn (Genetic Programming for Symbolic Regression) \cite{stephens2015gplearn}. These algorithms employ different techniques to discover symbolic expressions from data, and their processes are very similar to each other.

They are SR libraries that combine genetic programming with machine learning techniques to discover mathematical expressions. The first step of their processes is creating an initial population of candidate equations represented by mathematical expressions composed of simple mathematical operations ($+,-,\times,\div$), functions (e.g., sine, cosine, exponential), and variables. Subsequently, each candidate equation is evaluated against the given dataset, and its performance is assessed using a fitness function, that measures how well the equation fits the data, typically by calculating the mean squared error (MSE) or a similar metric. A genetic algorithm is used to select the best-performing candidate equations for the next generation. Equations that fit the data well are more likely to be selected, while less fit equations may be removed. Genetic operations like crossover (combining parts of two equations) and mutation (making small changes to an equation) are applied to the selected equations to create a new generation of candidate equations. This process iterates through multiple generations, continually improving the equations' fitness until a termination condition, such as a maximum number of generations, or a threshold fitness level, is met.

\section{Results}\label{sec:results}

In this section, the results of our simulations are reported and discussed. The first two subsections \ref{subsec:PK} and \ref{subsec:GI} show the performance of X-TFC and PINNs in parameter discovery and gray-box identification for both Pharmacokinetics and Ultradian Endocrine model. The outputs of the gray-box identification are used as input in the symbolic regression algorithms for the symbolic distillation of both NN-based methods, whose results and performance are shown in subsections \ref{subsec:SR_PK} and \ref{subsec:SR_GI}.


\subsection{Pharmacokinetics}\label{subsec:PK}

In the parameter discovery test case, we aim to infer the value of the parameters $k_g = 0.72 h^{-1}$ and $k_b = 0.15 h^{-1}$ of the system of ODEs in Eqs. \eqref{eq:PK}, given a certain numbers of available data points of $B$, $G$, and $U$. The results and performance for both X-TFC and PINNs are reported in Table \ref{tab:PK_param}, simulating the variation of drug concentration in the three compartments for a time domain of 50 hours. The number of data points used varies from 10 to 100, and both methods show great accuracy in retrieving both the parameters governing the ODEs. The accuracy of the methods is evaluated with the absolute difference between the nominal value of the parameters and their inference. As expected, we see an increase of accuracy while increasing the number of data points, but one can see that both methods can give great precision even for meager dataset (10 data points -- one every five hours).

For the pharmacokinetics inverse problem, in PINNs we utilized the Adam optimization with $N_c = 100$, learning rate ($lr$) of $1\times 10^{-4}$, and we conducted training for 30,000 iterations. Notably, in this context, the application of self-adaptive loss balancing weights was deemed unnecessary, and the two-phase training method was not employed. We perform the computational experiments for PINNs on NVIDIA's GeForce RTX 3090 GPUs, which are powered by NVIDIA’s 2nd generation RTX Ampere architecture. The GPU has 10496 core and endowed with 24 GB of GDDR-6X memory.

Since X-TFC uses a domain decomposition technique, we report the number of iterations needed from the iterative least-squares for each sub-domain, with an iteration tolerance set equal to 1e-09. When a decomposition of the domain is required to increase the accuracy of the results (cases with 20, 50, and 100 data points), the inferred parameter is given by the mean of the parameters inferred in each subdomain. The X-TFC results reported in Table \ref{tab:PK_f} are obtained with certain neural networks hyperparameters setup. The tuning hyperparameters are $N$ number of points per sub-domain, $L$ number of neurons, and $t_{step}$ the length of each subdomain. These setups for each simulation are reported in Table \ref{tab:PK_XTFC_hyperparam}, made with a Intel(R) Xeon(R) W-2255 CPU @ 3.70GHz machine.
\begin{table}[h!]
\centering
\begin{tabular}{lcccc}
\hline
 \multicolumn{5}{c}{\textbf{X-TFC}}   \\ \hline
\multicolumn{1}{c}{\textbf{\# data}} & \multicolumn{2}{c}{\textbf{absolute error}}  & \multicolumn{1}{c}{\textbf{\# of}} & \multicolumn{1}{c}{\textbf{comp.}} \\ 
\multicolumn{1}{c}{\textbf{points}} & \textbf{k\textsubscript{g}} & \textbf{k\textsubscript{b}}  & \multicolumn{1}{c}{\textbf{iter.}}  & \textbf{time [s]}  \\ \hline \hline
10 & 4.60e-03 & 1.05e-04    & 4         & 0.03    \\
20 & 1.65e-03 & 6.57e-05    & 37 & 0.05    \\
50 & 1.58e-04 & 5.79e-06    & 12 & 0.05  \\
100 & 3.54e-06 & 2.26e-07   & 6 & 0.05 \\
\hline \hline
  \multicolumn{5}{c}{\textbf{PINNs}} \\ \hline
\multicolumn{1}{c}{\textbf{\# data}} & \multicolumn{2}{c}{\textbf{absolute error}}  & \multicolumn{1}{c}{\textbf{\# of}} & \multicolumn{1}{c}{\textbf{comp.}} \\ 
\multicolumn{1}{c}{\textbf{points}} & \textbf{k\textsubscript{g}} & \textbf{k\textsubscript{b}}  & \multicolumn{1}{c}{\textbf{iter.}}  & \textbf{time [s]} \\ \hline \hline
10 & 1.90e-02   &  4.43e-05 & 3e04 & 37.32    \\
20  & 3.94e-06 &  5.81e-06 &3e04  & 36.00  \\
50 & 1.04e-05  &   5.47e-06 &3e04  & 35.51 \\
100  & 1.04e-05  &  5.47e-06 &3e04 & 36.47 \\ \hline
\end{tabular}
\caption{Pharmacokinetics model: performance of X-TFC and PINNs for parameter discovery for time range [0,50] hours. Refer to Table 1 for X-TFC hyperparameters.}\label{tab:PK_param}
\end{table}\\
GPUs, renowned for their inherently parallel architecture, excel in efficiently distributing specific computations across a multitude of cores. As the volume of data points grows, the potential for enhanced parallelization efficiency becomes evident, potentially resulting in reduced computation times. It is worth highlighting that when employing GPUs, computational times may decrease as the number of data points increases, as illustrated in the table \ref{tab:PK_param} depicting the results of the PINNs method. This phenomenon is particularly noticeable due to our utilization of GPUs for this method.

\begin{figure}
    \centering
    \begin{subfigure}[b]{0.7\linewidth}
        \includegraphics[width=\linewidth]{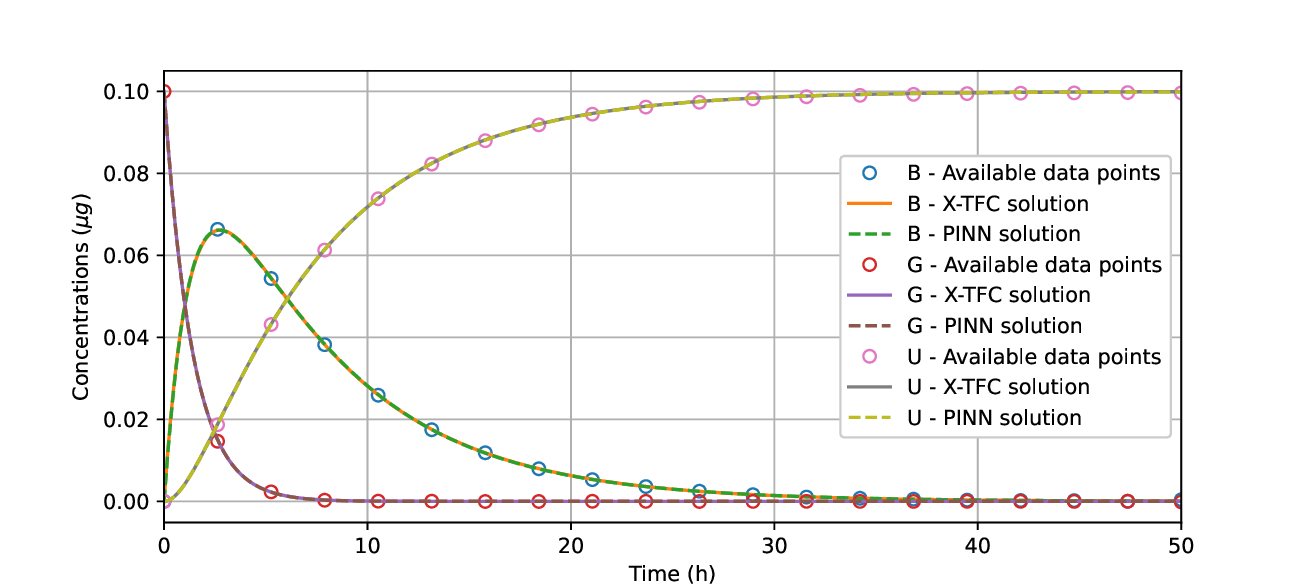}
        \caption{State variables $B$, $G$, and $U$ comparison.}
        \label{fig:var_drug}
    \end{subfigure}
    \hfill
    \begin{subfigure}[b]{0.7\linewidth}
        \includegraphics[width=\linewidth]{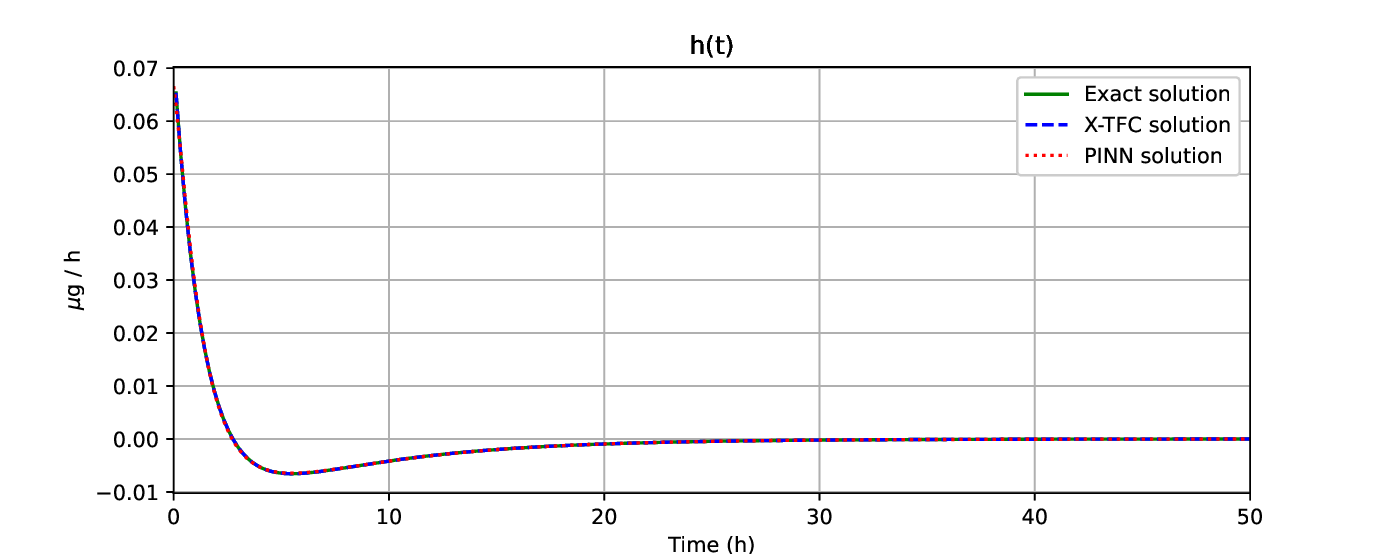}
        \caption{Unknown term $h(t)$ comparison.}
        \label{fig:f_t_drug}
    \end{subfigure}
    \caption{Pharmacokinetics model: comparison between exact solution vs. X-TFC and PINNs solutions, for (a) the variables $B$, $G$, and $U$, with 20 data points per variable, and for (b) for the unknown term $h(t)$.}
\end{figure}

In the gray-box identification test case for the Pharmacokinetics model, we aim to obtain the right-hand-side unknown term $h(t)$ of the first ODE of the system \eqref{eq:pharmacok_f}. X-TFC and PINNs results and performance for a simulation of 50 hours are shown in Table \ref{tab:PK_f}. Performance are evaluated via Mean Absolute Error (MAE):
\begin{equation*}
    MAE = \frac{\sum_{i=1}^N |\hat{h}_i(t) - h_i(t)|}{N} ,
\end{equation*}
Root Mean Squared Error (RMSE):
\begin{equation*}
    RMSE = \sqrt{ \frac{ \sum_{i=1}^N ( \hat{h}_i(t) - h_i(t)  )^2}{N}   } ,
\end{equation*}
and Relative Error (RE):
\begin{equation*}
    RE = \dfrac{\sqrt{ \sum_{i=1}^N (\hat{h}_i(t) - h_i(t))^2  }}{\sqrt{ \sum_{i=1}^N \hat{h}_i(t)^2 }}
\end{equation*}
where $\hat{h}(t)$ and $h(t)$ are the exact and learned solutions, respectively. Also, for these test cases, we can see how both methods can perform a good inversion of the unknown term $h(t)$ given few data samples. Figure \ref{fig:var_drug} shows the learned concentrations in time of the three state variable  $B$, $G$, and $U$ for X-TFC and PINNs solutions vs. the exact solution (given by 50 data points), while the learned function $h(t)$ is plotted in Figure \ref{fig:f_t_drug}.
\begin{table}[h!]
\centering
\begin{tabular}{lccccc}
\hline
\multicolumn{1}{c}{} &  \multicolumn{5}{c}{\textbf{X-TFC}}  \\ \hline
\multicolumn{1}{c}{\textbf{\# data}}  & \multicolumn{3}{c}{\bf{$h(t)$}} & \multicolumn{1}{c}{\textbf{\# of}} & \multicolumn{1}{c}{\textbf{comp. time}}  \\ 
\multicolumn{1}{c}{\textbf{points}}  & \multicolumn{1}{c}{\bf{MAE}}  & \multicolumn{1}{c}{\bf{RMSE}}  & \multicolumn{1}{c}{\bf{RE}} & \multicolumn{1}{c}{\textbf{iter.}} &  [sec.]    \\ \hline \hline
10 &  3.57e-03  & 6.91e-03 & 3.00e-01 & 3  &  0.015 \\
20 &  1.26e-04    & 4.61e-04 & 2.91e-02 & 3,2 & 0.015   \\
50 & 3.87e-07 & 6.39e-07 &  5.75e-05   & 3,3  & 0.05      \\
100  & 1.56e-08  & 9.37e-08 & 9.97e-06 & 3,3,2,2  & 0.08    
\\ \hline \hline
\multicolumn{1}{c}{} &  \multicolumn{5}{c}{\textbf{PINNs}}  \\ \hline
\multicolumn{1}{c}{\textbf{\# data}}  & \multicolumn{3}{c}{\bf{$h(t)$}} & \multicolumn{1}{c}{\textbf{\# of}} & \multicolumn{1}{c}{\textbf{comp. time}}  \\ 
\multicolumn{1}{c}{\textbf{points}}  & \multicolumn{1}{c}{\bf{MAE}}  & \multicolumn{1}{c}{\bf{RMSE}}  & \multicolumn{1}{c}{\bf{RE}} & \multicolumn{1}{c}{\textbf{iter.}} &  [sec.]  \\ \hline \hline
10 &  1.26e-04  & 5.57e-04  &7.10e-02  &3e04, 1e02 & 141.68  \\
20 &   1.09e-04  & 4.82e-04  & 6.01e-02 &3e04, 1e02 & 145.97  \\
50 & 6.59e-05  & 2.26e-04  & 2.78e-02  &3e04, 1e02& 140.81   \\
100    & 6.54e-05  & 1.84e-04  & 2.26e-02  & 3e04, 1e02& 143.37   
\\ \hline
\end{tabular}
\captionsetup{justification=raggedright}
\caption{Pharmacokinetics model: Unknown term discovery for time range [0,50] hours. Comparison between X-TFC and PINNs performance via $MAE$, $RMSE$, $RE$, and computational time for different number of data points. The initial number in the '\# of Iter.' column for PINNs represents the iterations during the primary training stages using Adam optimization while the second number corresponds to the training stage utilizing L-BFGS.}\label{tab:PK_f}
\end{table}

As presented in Table \ref{tab:PK_param} and \ref{tab:PK_f}, our comparative analysis reveals valuable insights into the performance of the X-TFC and PINNs methods when applied to the same problem with varying data sizes within the same time range. For smaller size of the dataset (e.g., 10 data points), the PINNs method can achieve better performance in accuracy, especially for the gray-box test case, showing its inherent performance in handling sparse datasets for approximating complex functions, due to the high expressivity of the deep neural network. Conversely, as the dataset size increases, the performance of X-TFC method in terms of accuracy improves substantially. Its computational speed, a distinct advantage, allows it to effectively capitalize on larger datasets. With more data points, the X-TFC method can produce increasingly accurate results, eventually surpassing the accuracy achieved by the PINNs method.  Despite the initial accuracy advantage of PINNs, it reaches a point where further increasing the dataset size does not significantly improves accuracy with the same setup, while still keeping great performance. This is probably due to the optiization error, and overcoming this limitation may involve architectural enhancements, such as increasing the neural network's depth, employing different optimization algorithms, or implementing alternative techniques. In contrast, the X-TFC method continues to benefit from additional data, showcasing its scalability and adaptability. In summary, for problems with small datasets, the PINNs method excels in providing accurate solutions. For larger datasets the X-TFC method becomes increasingly competitive, offering the potential for superior accuracy with adequate computational resources.\\
\begin{table}[h!]
\centering
\begin{tabular}{lccc}
\hline
\multicolumn{4}{c}{\textbf{X-TFC}} \\
\hline \hline
\multicolumn{1}{c}{\textbf{\# data points}} & \textbf{N} & \textbf{L}  & \multicolumn{1}{c}{$\boldsymbol{t_{step}}$}  \\ \hline \hline
10 & 10  &  100   &  50 \\
20 &  11 &  100   &  25  \\
50 &  26 &  100    & 25  \\
100 & 26  &  100   &  12.5  \\
\hline
\end{tabular}
\caption{Pharmacokinetics model: X-TFC hyperparameters setup for parameter discovery and unknown term discovery, for time range [0,50] hours.}\label{tab:PK_XTFC_hyperparam}
\end{table}

\begin{table}[h!]
\centering
\begin{tabular}{lc}
\hline
\multicolumn{2}{c}{\textbf{PINNs parameters}} \\
\hline \hline
Optimizer & Adam, LBFGS \\
Activation Function & Tanh \\
Number of Iterations & 5000, 25000, 100\\
Architecture of main NN & 50, 7 \\
Architecture of second NN & 20, 5 \\
Learning Rate for main NN & 0.001 \\
Learning Rate for second NN & 0.0001 \\
Number of Collocation Points & 500 \\
\hline
\end{tabular}
\caption{Pharmacokinetics model: PINNs parameters setup for the discovery of unknown terms in the  over a time range of [0,50] hours. The initial and second numbers in the 'Number of Iterations' Row represent the iterations during the primary and secondary training stages using Adam optimization. The third number corresponds to the training stage utilizing L-BFGS. The first and second numbers in the 'Architecture of Neural Networks' indicate the width and depth, respectively.}\label{tab:PINN_drug_hyperparam}
\end{table}



\newpage

\subsection{Ultradian Endocrine model}\label{subsec:GI}

\begin{figure}[h!]
    \centering
    \begin{subfigure}[b]{0.9\textwidth}
        \includegraphics[width=\linewidth]{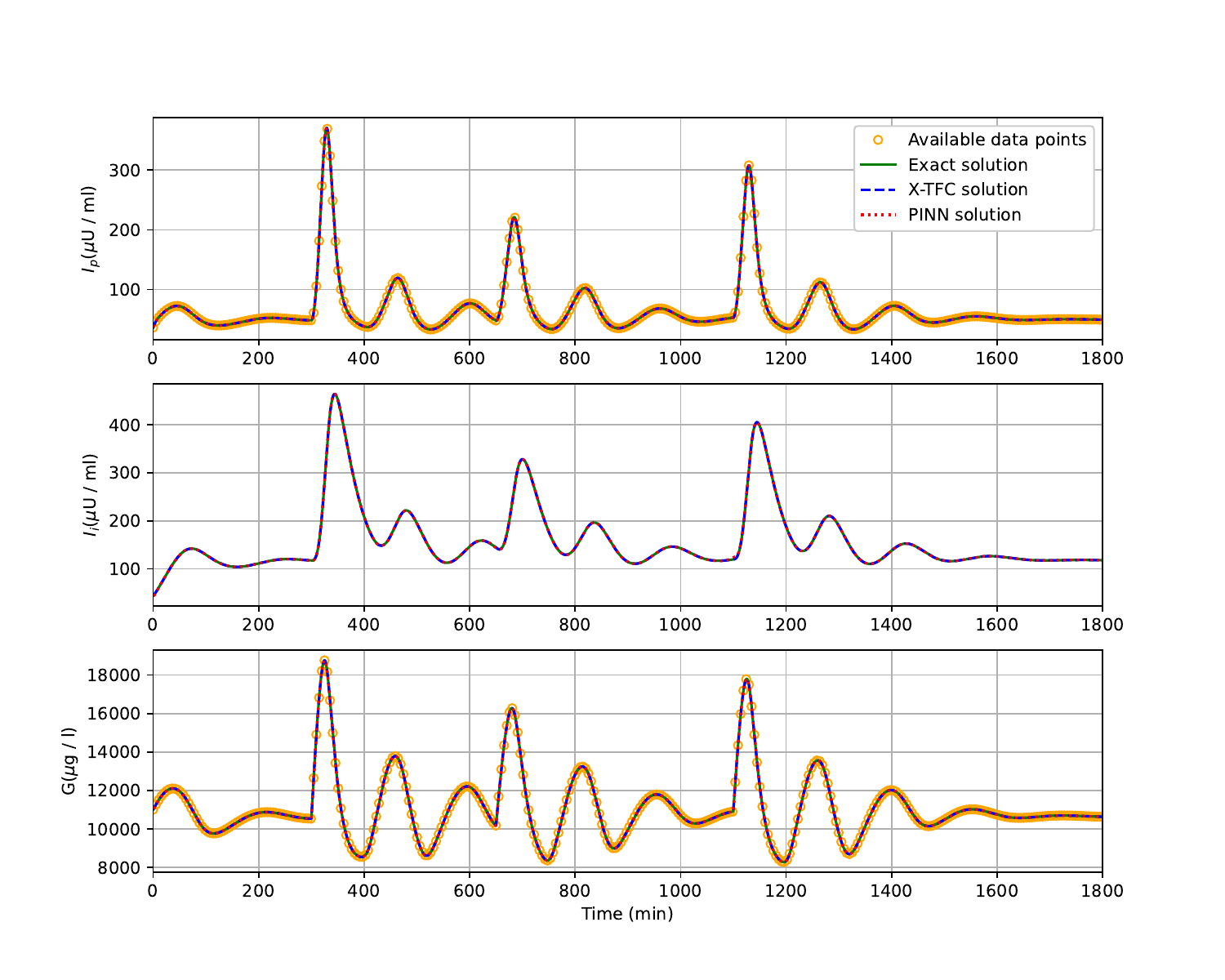}
        \caption{Ultradian Endocrine model: state variables $I_p$, $I_i$, and $G$: 360 observed data points of $I_p$ and $G$, exact solution, and learned solutions via X-TFC and PINNs methods.}
        \label{fig:var_glucose}
    \end{subfigure}
    \begin{subfigure}[b]{0.9\textwidth}
        \includegraphics[width=\linewidth]{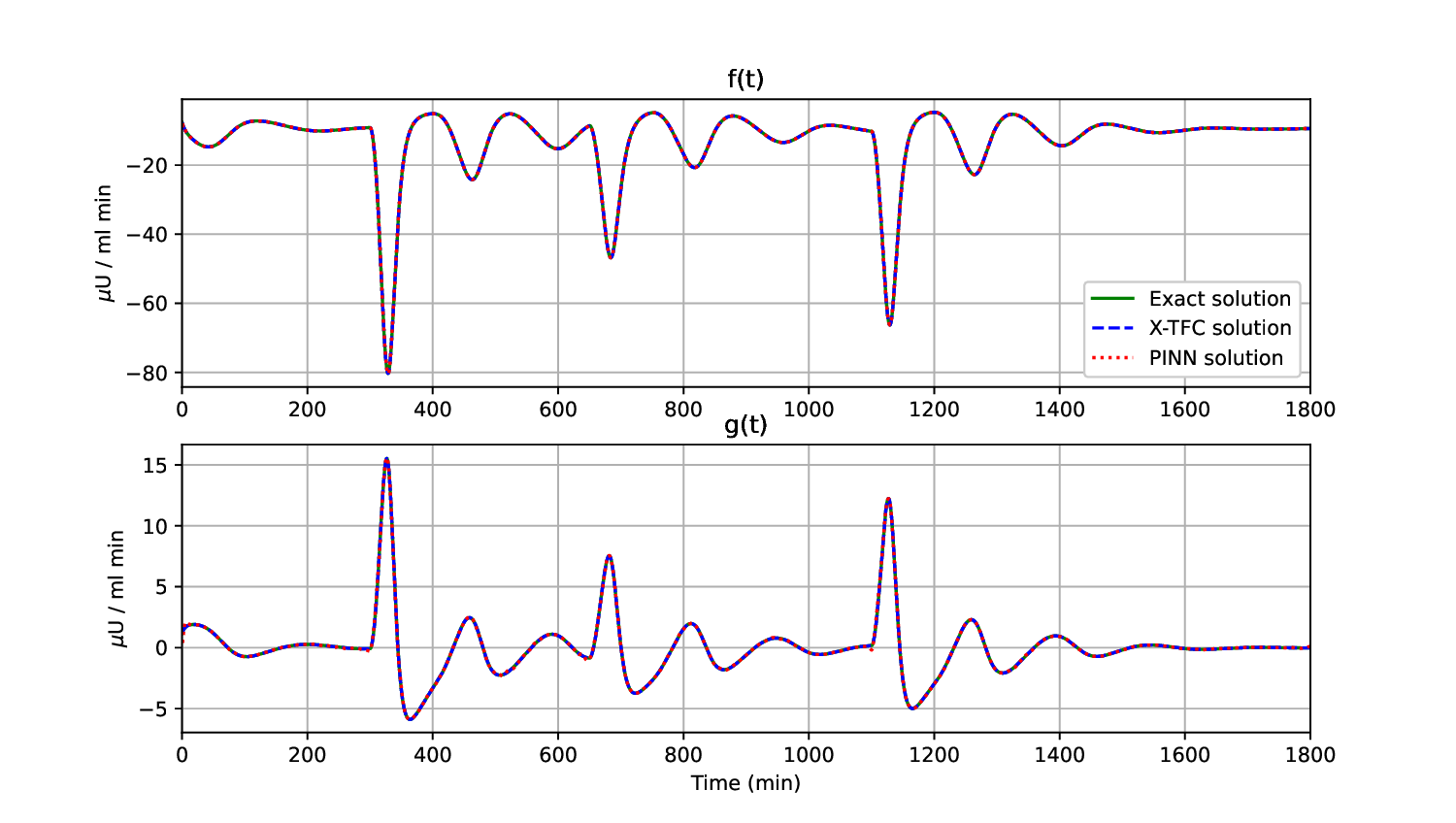}
        \caption{Ultradian Endocrine model: unknown terms $f(t)$ and $g(t)$: exact solution, and learned solutions via X-TFC and PINNs methods.}
        \label{fig:f_t_glucose}
    \end{subfigure}
    \caption{Glucose-insulin interaction model: comparison between exact solution vs. X-TFC and PINNs solutions for: (a) the variables $I_p$, $I_i$, and $G$ (top to bottom), and (b) unknown terms $f(t)$ and $g(t)$ (top to bottom).}
\end{figure}

The results of the parameter discovery test case for the Ultradian Endocrine model are reported in Table \ref{tab:GI_param}, as the absolute difference between the nominal and inferred values of the parameters. Our simulations were conducted for the discovery of five parameters. However, the PINNs algorithm proved to be very effective in system identification, discovering up to 21 parameters of the ultradian endocrine model using only data for $G$ and $I_p$. As presented in \cite{daneker2023systems}, using only 360 data points for G, the PINNs algorithm was able to discover 17 parameters accurately, which is challenging and not possible for the X-TFC algorithm to do with a small amount of data on only one state variable.\\
With X-TFC, we are able to retrieve the parameters already in the first sub-domain, thus further iteration of the algorithm would be redundant, allowing us to further speed up the computation. In the context of PINNs, the obtained results are contingent on the learning process. Notably, the neural network's capacity to learn effectively is closely tied to the temporal scope of the problem. Specifically, the neural network may not yield accurate approximations within a smaller time range, which corresponds to a reduced dataset size.\\  
\begin{table}[h!]
\centering
\begin{tabular}{lccccc|cc}
\hline
\multicolumn{7}{c}{\textbf{X-TFC}} \\ \hline
\textbf{\# data} & \multicolumn{5}{c|}{\textbf{absolute error}}  & \multicolumn{1}{c}{\textbf{\# of}}   & \multicolumn{1}{c}{\textbf{comp.}}\\ 
\textbf{points} & \bf{$E$} & \bf{$t_p$} & \bf{$t_i$} & \bf{$R_m$} & \bf{$a_1$}  & \textbf{iter.}  & \textbf{time [s]}  \\ \hline \hline
360   & 3.07e-03  &  2.44e-01 &  8.29e-01   &   1.10e-01 &  2.76e-02 & 27  &  0.7  \\
450   & 6.36e-06  &  7.01e-03 &  5.75e-03 &  2.13e-00 &  1.00e-02 &  31 &   0.8 \\
600   &  5.43e-07 &  4.10e-03 &  6.62e-04 &  1.71e-00  &  8.21e-03 &  61 &   1.5 \\
900   &  1.48e-07 & 1.79e-06   & 1.26e-04  &  9.39e-04 & 4.63e-06  &  84  &  1.5 \\
1800    & 1.03e-10   &  4.17e-09  & 8.78e-08  &  1.28e-06   &  6.05e-09 & 105  &  2.5  \\   
   \hline \hline
\multicolumn{7}{c}{\textbf{PINNs}} \\ \hline
\textbf{\# data} & \multicolumn{5}{c|}{\textbf{absolute error}} & \multicolumn{1}{c}{\textbf{\# of}}  & \multicolumn{1}{c}{\textbf{comp.}}\\ 
\textbf{points} & \bf{$E$} & \bf{$t_p$} & \bf{$t_i$} & \bf{$R_m$} & \bf{$a_1$}   & \textbf{iter.}  & \textbf{time [s]}  \\ \hline \hline
360   & 7.54e-05 & 1.46e-02& 3.97e-01& 5.64e-01& 6.32e-04 & 6e05 & 2494.6  \\
450   & 4.01e-05 & 2.20e-04& 3.71e-03& 1.01e-02& 1.80e-04 & 6e05 & 2455.2  \\
600   &  1.01e-05& 6.06e-04& 3.78e-03& 2.14e-02& 1.42e-04 & 6e05 & 2577.1  \\
900   &  3.06e-05& 1.03e-04& 1.25e-02& 7.64e-03& 1.43e-04 & 6e05 & 2631.6  \\
1800  & 1.80e-05 & 6.92e-04& 9.92e-04& 2.28e-02& 1.11e-04 & 6e06 & 2946.5  \\  
 \hline
\end{tabular}
\caption{Ultradian Endocrine model: parameter discovery via X-TFC and PINNs algorithms. The performance of the two methods is given by the absolute difference between nominal values and inferred values. On the right we also present computational times in seconds.}\label{tab:GI_param}
\end{table}
In the gray-box identification case, we aim to infer the two unknown terms $f(t)$ and $g(t)$ in the system of ODEs \eqref{eq:GI_system_f_t}, from available data of the variables $I_p$ and $G$. In Table \ref{tab:GI_f_t}, the $MAE$, $RMSE$, $RE$, and computational times are reported for both X-TFC and PINNs frameworks, for different amount of data points, from 360 to 1800 (i.e., data available every 5, 4, 3, 2, and 1 minutes), in a simulation of 1800 minutes. For X-TFC, a domain decomposition of several subdomains is needed, thus the number of iterations reported in the table refers to the average number of iterations in one subdomain. The hyperparameters for the X-TFC neural networks, as well as the configuration of parameters for the PINNs, employed to generate the results presented in Table \ref{tab:GI_f_t}, are documented in Tables \ref{tab:GI_XTFC_hyperparam} and \ref{tab:PINN_GI_hyperparam}, respectively. The first three state variables of the model learned by X-TFC and PINNs are plotted vs. the exact solution in Figure \ref{fig:var_glucose}, while the two learned functions $f(t)$ and $g(t)$ are plotted in Figure \ref{fig:f_t_glucose}. In both figures, the overlap of the solutions of both frameworks is clear. 

As evidenced by the data presented in Tables \ref{tab:GI_param} and \ref{tab:GI_f_t}, encompassing both gray-box and inverse problem scenarios, and spanning across both this model and the pharmacokinetics model, a discernible pattern emerges concerning the impact of dataset size on method performance.

In the case of the X-TFC method, an increase in the number of data points leads to progressively more accurate results. However, it is noteworthy that when confronted with a relatively small dataset, the PINNs method exhibits superior performance, characterized by heightened accuracy and reduced absolute error. For instance, in Table \ref{tab:GI_f_t}, the PINNs method demonstrates better efficacy with merely 360 and 450 data points. Nevertheless, as the dataset size grows, the X-TFC method surpasses PINNs in both accuracy and computational efficiency.

In summary, the choice between the X-TFC and PINNs methods should be made judiciously, with careful consideration of dataset size and noise levels. While the X-TFC method excels with larger datasets, the PINNs method exhibits a unique strength in scenarios involving smaller datasets or noisy data, where it achieves greater accuracy.

\begin{table}[h!]
\centering
\begin{tabular}{lccc|ccc|cc}
\hline
 \multicolumn{9}{c}{\textbf{X-TFC}} \\ \hline
\multicolumn{1}{c}{\textbf{\# data}}  & \multicolumn{3}{c|}{\bf{$f(t)$}}   & \multicolumn{3}{c|}{\bf{$g(t)$}} & \multicolumn{1}{c}{\textbf{\# of}} & \multicolumn{1}{c}{\textbf{comp. time}} \\ 
\multicolumn{1}{c}{\textbf{points}}  & \multicolumn{1}{c}{\bf{MAE}}  & \multicolumn{1}{c}{\bf{RMSE}}  & \multicolumn{1}{c|}{\bf{RE}}  & \multicolumn{1}{c}{\bf{MAE}}  & \multicolumn{1}{c}{\bf{RMSE}}  & \multicolumn{1}{c|}{\bf{RE}} & \multicolumn{1}{c}{\textbf{iter.}}  &  [sec.]  \\ \hline \hline
360 & 7.08e-02  & 3.14e-01  & 1.93e-02   &    1.33e-01   & 4.14e-01   &  1.67e-01  & 3 & 0.15  \\
450 & 6.83e-03 & 3.31e-02 &  2.03e-03 & 6.45e-02 & 2.76e-01 & 1.11e-01 & 3  & 0.20 \\
600 &  2.62e-03 &  1.10e-02  &  6.70e-04  &  2.58e-02  & 1.25e-01  & 4.92e-02  &  3  &  0.20 \\
900 &  7.96e-04 & 2.74e-03    &  1.69e-04  &  2.46e-02  & 6.46e-02  &  2.61e-02 &  3  &  0.25 \\
1800 &  1.65e-04  &  6.96e-04  & 4.19e-05   & 2.04e-03    & 5.72e-03   & 2.23e-03  &   3  &  0.35  \\ \hline \hline
\multicolumn{9}{c}{\textbf{PINNs}} \\ \hline
\textbf{\# data} & \multicolumn{3}{c|}{\textbf{$f(t)$}} & \multicolumn{3}{c|}{\textbf{$g(t)$}} & \textbf{\# of} & \textbf{comp. time} \\ 
\textbf{points} & \textbf{MAE} & \textbf{RMSE} & \textbf{RE} & \textbf{MAE} & \textbf{RMSE} & \textbf{RE} & \textbf{iter.} & \textbf{[sec.]} \\
\hline \hline
360 & 1.99e-02 & 8.42e-02 & 5.07e-03 & 5.49e-02 & 1.32e-01 & 5.12e-02 & 1e06 & 3883.68 \\
450 & 1.57e-02 & 7.58e-02 & 4.56e-03 & 4.04e-02 & 8.61e-02 & 3.35e-02 & 1e06 & 3958.43 \\
600 & 8.57e-03 & 6.99e-02 & 4.21e-03 & 3.58e-02 & 9.18e-02 & 3.58e-02 & 1e06 & 4028.44 \\
900 & 8.27e-03 & 3.80e-02 & 2.29e-03 & 3.77e-02 & 7.77e-02 & 3.03e-02 & 1e06 & 4177.49 \\
1800 & 7.89e-03 & 5.99e-02 & 3.61e-03 & 3.14e-02 & 6.02e-02 & 2.35e-02 & 1e06 & 4917.81 \\
\hline
\end{tabular}
\caption{Ultradian Endocrine model: unknown terms discovery for time range [0,1800] minutes. X-TFC and PINNs performance in terms of $MAE$, $RMSE$, $RE$, number of iterations, and computational time for different numbers of data points.}
\label{tab:GI_f_t}
\end{table}

\begin{table}[h!]
\centering
\begin{tabular}{lccc}
\hline

\multicolumn{1}{l}{\textbf{\# data points}} & \textbf{N} & \textbf{L}  & \multicolumn{1}{c}{$\boldsymbol{t_{step}}$}  \\ \hline \hline
360 & 4  &  5   &  15 \\
450 &  5 &  10   &  16  \\
600 &  5 &  15    & 12  \\
900 & 5  &  20   &  8  \\
1800 & 6  &  20   &  5  \\
\hline
\end{tabular}
\caption{Ultradian Endocrine model: X-TFC hyperparameters setup for parameter discovery and unknown terms discovery, for time range [0,1800] minutes.}\label{tab:GI_XTFC_hyperparam}
\end{table}

\begin{table}[h!]
\centering
\begin{tabular}{lc}
\hline
\multicolumn{2}{c}{\textbf{PINNs parameters}} \\
\hline \hline
Optimizer & Adam \\
Activation Function & Swish \\
Number of iterations & 10000, 1000000 \\
Architecture of main NN & 128, 4 \\
Architecture of second NN & 32, 4 \\
Learning Rate for main NN & 0.001 \\
Learning Rate for second NN & 0.001 \\
Number of Collocation Points & 1800 \\
\hline
\end{tabular}
\caption{Ultradian Endocrine model: PINNs parameters setup for unknown terms discovery, for the time range [0,1800] minutes. The first and second numbers in the 'Architecture of Neural Networks' indicate the width and depth, respectively. The initial and second numbers in the ’Number of Iterations’ Row represent the iterations during the primary and secondary training stages.
}\label{tab:PINN_GI_hyperparam}
\end{table}

\newpage

\subsection{Symbolic Distillation of gray-box models recovered from X-TFC and PINNs methods}
After training of X-TFC and PINNs model, we obtain a gray-box models for $f(t),~g(t)$ and $h(t)$ parameterized by high dimensional parameters. Therefore, we perform symbolic regressions and fit a compact closed-form analytical expressions to $f(t), g(t)$ and $h(t)$ independently by using PySR \cite{cranmer2023interpretable} and gplearn \cite{stephens2015gplearn}. Both packages use a genetic algorithm to combine algebraic expressions stochastically.
The employed method shares similarities with the method of natural selection, as it assesses the ``fitness" of each expression based on its simplicity and accuracy. In this study, we consider binary operations  in the fitting process are $+,~-,$ and $~\times$. In symbolic regression, accuracy of recovered expressions is assessed through complexity, score, and loss. Complexity measures intricacy of the discovered equations in terms of the number of terms, mathematical operations, and the overall structure of the equations. Managing complexity is an important aspect of symbolic regression because overly complex equations can be difficult to interpret and may not generalize well to new data, leading to overfitting. Score in symbolic regression algorithm is typically used to  discover the mathematical expressions that maximize or minimize the chosen scoring metric while considering different combinations of mathematical operations and constants. Loss in symbolic regression typically refers to a mathematical function that quantifies the discrepancy between the predicted values generated by a symbolic expression or equation and the actual observed values in the dataset.

We represent the validation metrics for the model obtained from PySR with variation in loss and score against the complexity of symbolic expression.
The loss function can be considered as mean square error (MSE) or root mean square of error (RMSE) between actual and predicted outputs. However, the score is defined as the negative of the derivative of the log-loss with respect to the complexity. The complexity in PySR is defined as the number of nodes in an expression tree, irrespective of each node’s content. In the PySR implementation, we chose the candidate model with the highest score among expressions with a loss better than at least 1.5x the most accurate model represented by lower most loss function. In gplearn, we observe the variation of the loss function against length of the symbolic expression, and we choose the candidate model when complexity increases but the loss remains stagnant.

\subsubsection{Symbolic distillation of Pharmacokinetics  model}\label{subsec:SR_PK}
\begin{figure}[h!]
    \subfloat[Validation metrics for $h_\text{sym}$ using PySR]{{\includegraphics[width=8cm]{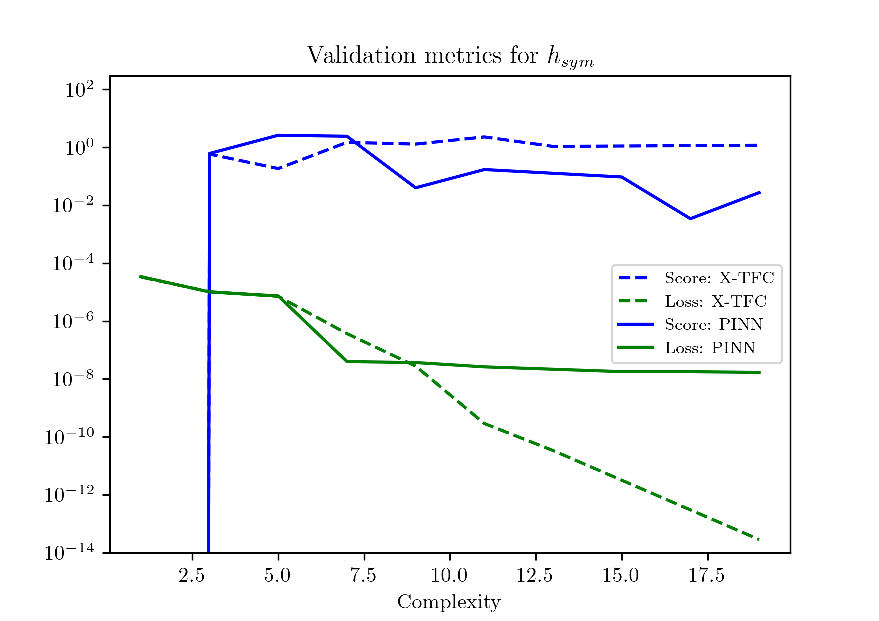} }}
     \subfloat[Validation metrics for $h_\text{sym}$ using gplearn]{{\includegraphics[width=8cm]{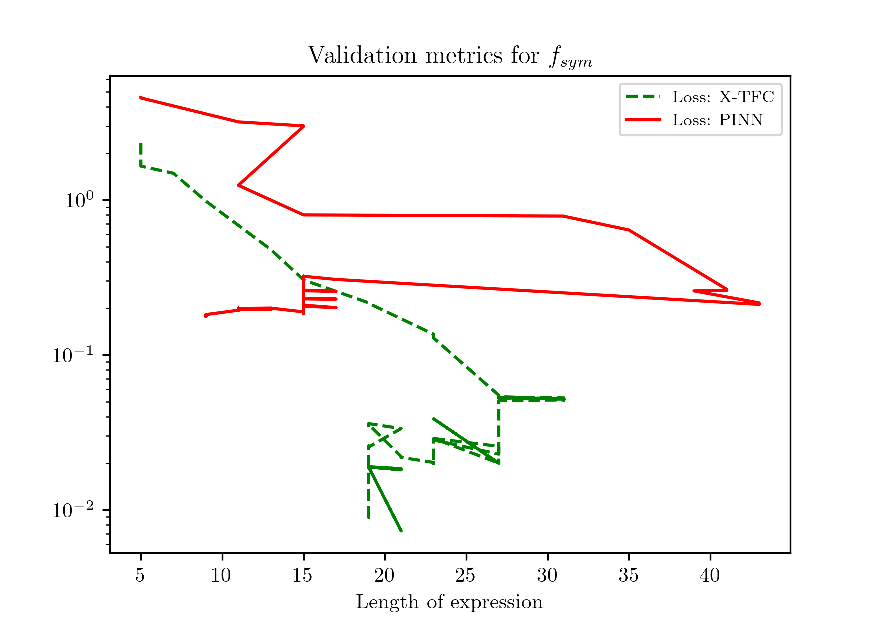} }}
     \caption{Pharmacokinetics  model: Validation metrics for the Pharmacokinetics model using for X-TFC and PINN based gray-box models. (a) represents variation in loss and score of symbolic models, obtained from PySR, with respect to complexity of expressions. Once convergence is achieved, the score remains constant as the complexity of the recovered expression increases and thus the criteria for selection of candidate symbolic with expression shown in \autoref{table:eq_PK_Model}. (b) represents variation in loss of symbolic models, obtained from gplearn, with respect to the length of expression. We choose the length of expression 9 and 19 for PINNs and X-TFC, respectively. These lengths of expressions corresponds to minimum loss for the regressed symbolic models with closed form expression shown in \autoref{table:eq_PK_Model}}
     \label{fig:sr_metric_PK}
\end{figure}

\begin{figure}[h!]
  \centering
  \includegraphics[width=0.45\textwidth]{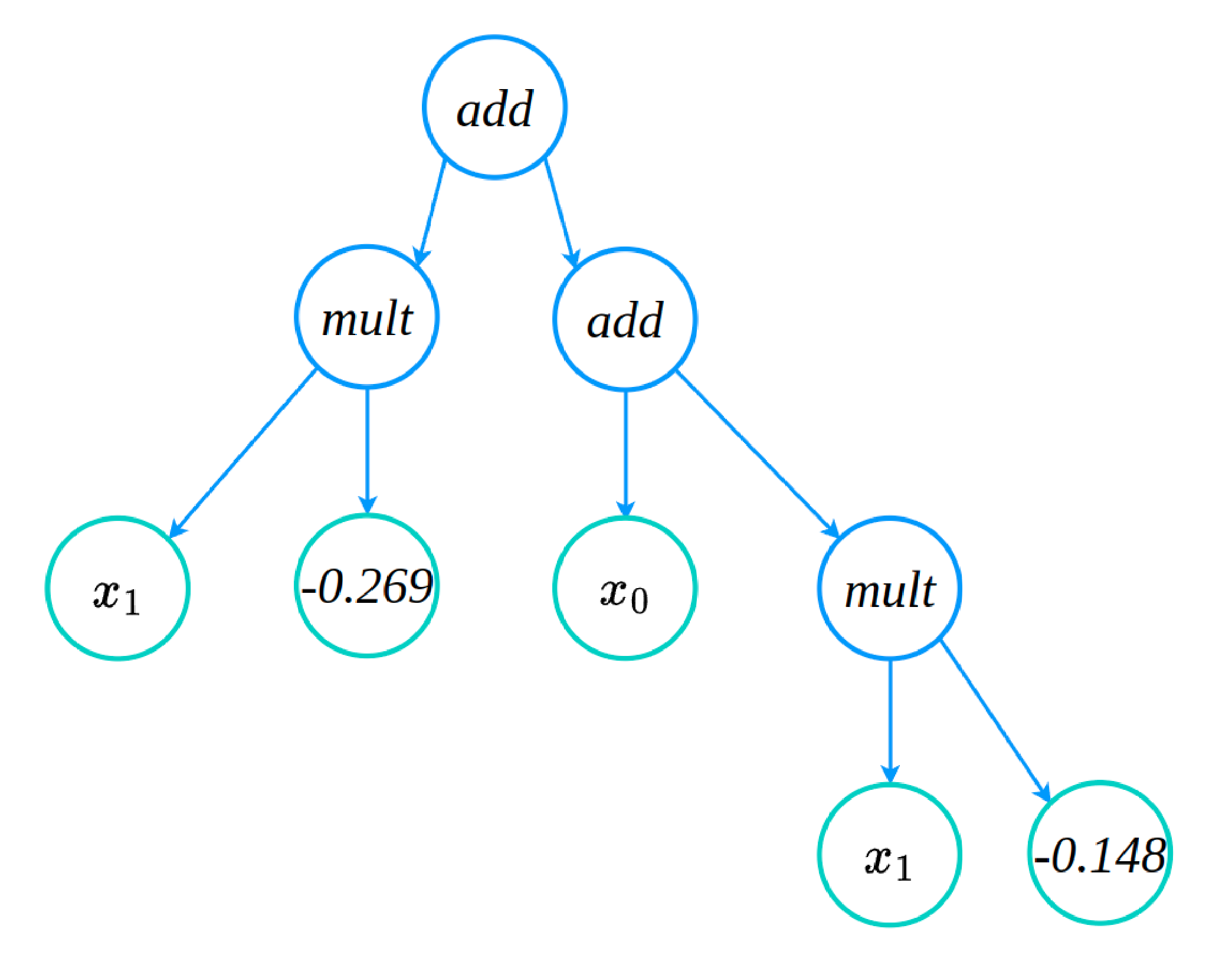}
  \caption{Pharmacokinetics  model: gplearn based evolved tree of binary operations in symbolic model recovered for black-box model $h_{sym}$ obtained from PINNs. It is to be noted that number of nodes in the tree corresponds to length of expressions, which is 9 for PINNs method. }
  \label{fig:symbolic_tree_pk}
\end{figure}

We perform symbolic regression for \eqref{eq: PINN_graybox}, in particular  \begin{align}\label{eq:pk_sr}
\frac{dB}{dt} = h_{sym}(G, B),
\end{align}
where we recover the expression $h_{sym}$ in terms of $G$ and $B$ using symbolic regressions.
\begin{table}[h!]
\centering
\begin{tabular}{lc|cc}
\hline
\multicolumn{1}{c}{}  & \multicolumn{1}{c|}{\bf{PySR}}   & \multicolumn{1}{c}{\bf{True Expression}} \\ 
\multicolumn{1}{c}{\textbf{Method}}  & \multicolumn{1}{c|}{\bf{$h_{sym}$}}  & \multicolumn{1}{c}{\bf{$h$}} 
\\ \hline \hline
 X-TFC &   $0.7199G - 0.15B$  & \multirow{2}{*}{$0.72G - 0.15B$}     \\
 PINNs  & $  0.7257G - 0.1559B$     &      
\\ \hline \hline
\multicolumn{1}{c}{}  & \multicolumn{1}{c|}{\bf{gplearn}}   & \multicolumn{1}{c}{\bf{True Expression}} \\ 
\multicolumn{1}{c}{\textbf{Method}}  & \multicolumn{1}{c|}{\bf{$h_{sym}$}}  & \multicolumn{1}{c}{\bf{$h$}} 
\\ \hline \hline
 X-TFC &   $0.7205G - 0.1507B$  & \multirow{2}{*}{$0.72G - 0.15B$}     \\
 PINNs  & $0.7310G - 0.1480B $     &      
\\ \hline
\end{tabular}
\caption{Pharmacokinetics model: Results of symbolic regression for gray-box identification using the PySR package (top) proposed by Cranmer \cite{cranmer2023interpretable} and the method implemented in gplearn \cite{stephens2015gplearn} package (bottom).}
\label{table:eq_PK_Model}
\end{table}
 In \autoref{table:eq_PK_Model}, we show the closed form symbolic models obtained from the packages PySR and gplearn for the black-box models recovered from X-TFC and PINNs approaches. From \autoref{table:eq_PK_Model} It is evident that symbolic  models are in very good agreement with the true models. Validation metrics for the models obtained from PySR and gplearn are shown in \autoref{fig:sr_metric_PK}. In \autoref{fig:sr_metric_PK}a we show the plots of loss and scores against the complexity of expressions for the  symbolic models obtained from  PySR. In \autoref{fig:sr_metric_PK}a it is evident that as complexity increases the scores remains constant for both the PINNs and X-TFC, which indicates convergence of candidate model. Similarly, the loss for PINN approach obtained the convergence very early, but the loss for X-TFC method keeps decreasing but complexity remains constant. Therefore, a candidate model with a complexity of 5 is appropriate and does not overfit. \autoref{fig:sr_metric_PK}b shows the validation metric of the symbolic model obtained from gplearn. Unlike PySR, gplearn provides the metric in terms of loss and length of expressions as population evolves. In \autoref{fig:sr_metric_PK}b, we plot the loss against the length of expression in symbolic models. The candidate models for PINN and X-TFC methods, shown in Table 2, correspond to length of 7 and 19, respectively. In \autoref{fig:symbolic_tree_pk}, we  show the evolved tree of binary operations, obtained from gplearn, in the symbolic model recovered for $h_{sym}$ obtained from PINNs. It is to be noted that the number of nodes (9) in \autoref{fig:symbolic_tree_pk} represents the the length of expression in the symbolic model.

\subsubsection{Symbolic Distillation of X-TFC and PINNs for Ultradian Endocrine model}\label{subsec:SR_GI}
\begin{figure}[h!]
    \subfloat[Validation metrics for $f_\text{sym}$]{{\includegraphics[width=8cm]{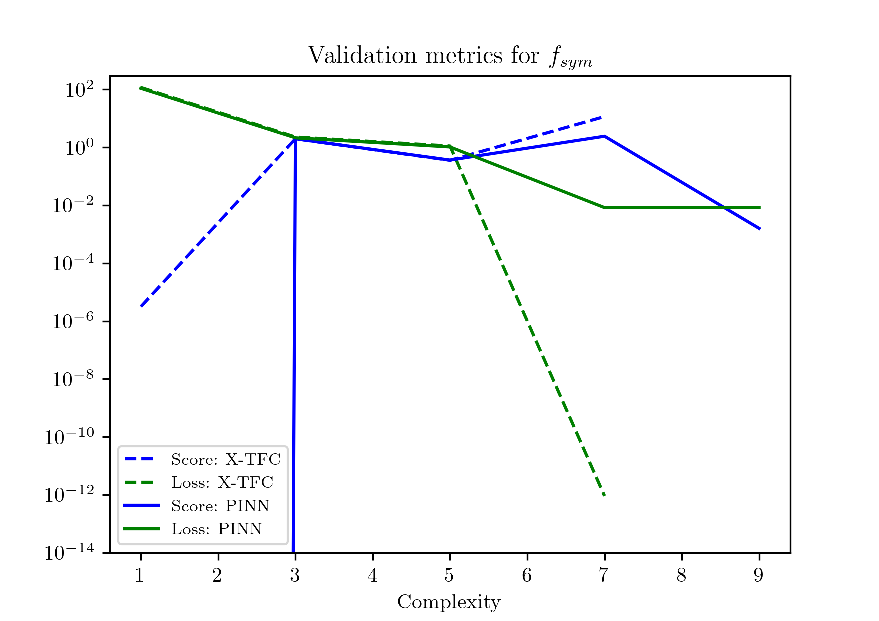} }}
     \subfloat[Validation metrics for $g_\text{sym}$]{{\includegraphics[width=8cm]{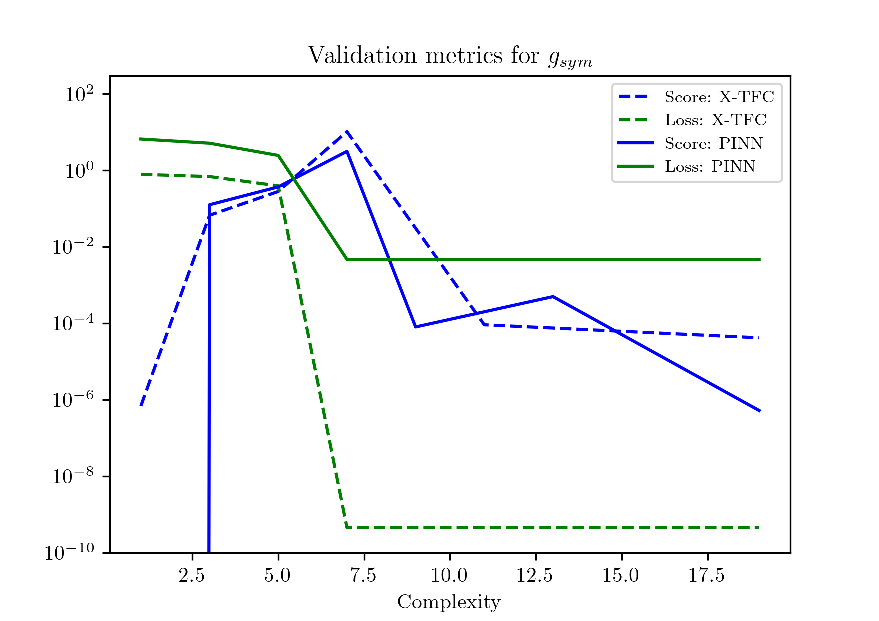} }}
     \caption{Ultradian Endocrine model: Validation metrics for PySR method. (a) $f_\text{sym}$ and (b) $g_\text{sym}$ are expressed by score and loss metrics against complexity of the expressions recovered using PySR. It is to be noted that, in both the plots once convergence is achieved, score remains unchanged as complexity increases.}
     \label{fig:sr_metric_GI_pysr}
\end{figure}

\begin{figure}[h!]
    \subfloat[Validation metrics for $f_\text{sym}$]{{\includegraphics[width=8cm]{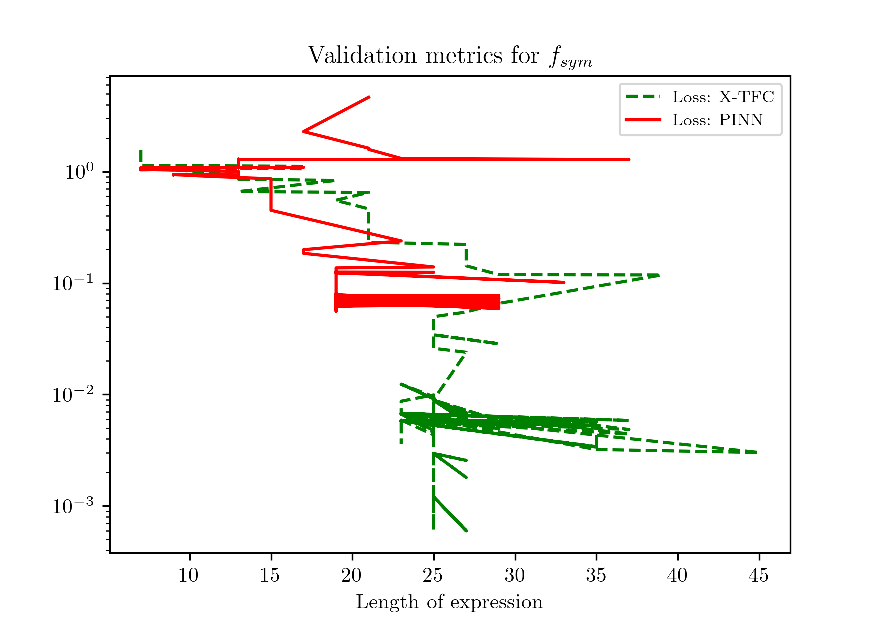} }}
     \subfloat[Validation metrics for $g_\text{sym}$]{{\includegraphics[width=8cm]{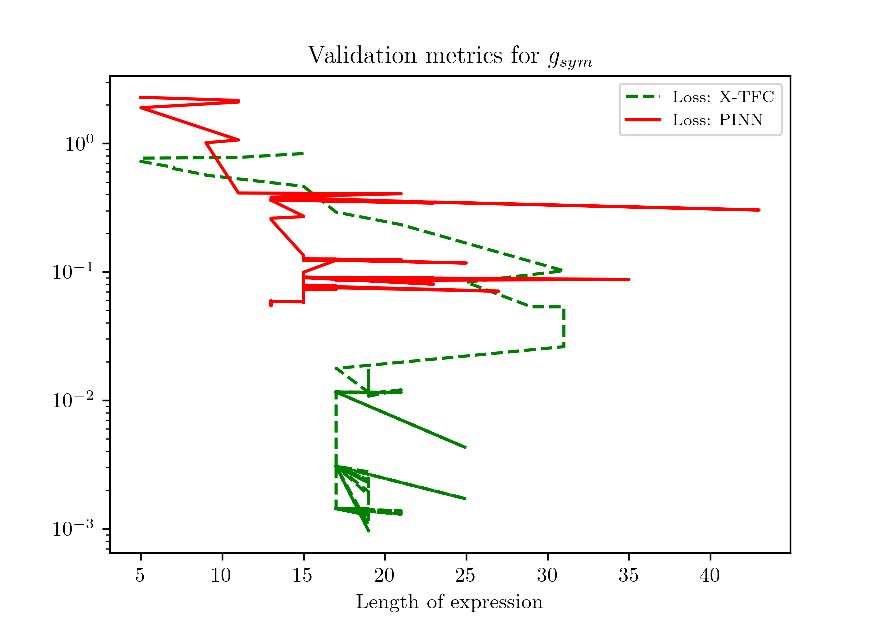} }}
     \caption{Ultradian Endocrine model: Validation metrics for gplearn method. (a) $f_\text{sym}$ and (b) $g_\text{sym}$ are expressed by MSE loss against length of the expressions recovered using gplearn and presented in \autoref{table:sbinn}. For $f_{sym}$, we choose length of expression 18 and 25 for PINNs and X-TFC, respectively. However, for $g_{sym}$, we choose length of expression 13 and 25 for PINNs and X-TFC, respectively.}
     \label{fig:sr_metric_GI_gplearn}
\end{figure}

\begin{figure}[h!]
  \centering
  \includegraphics[width=0.5\textwidth]{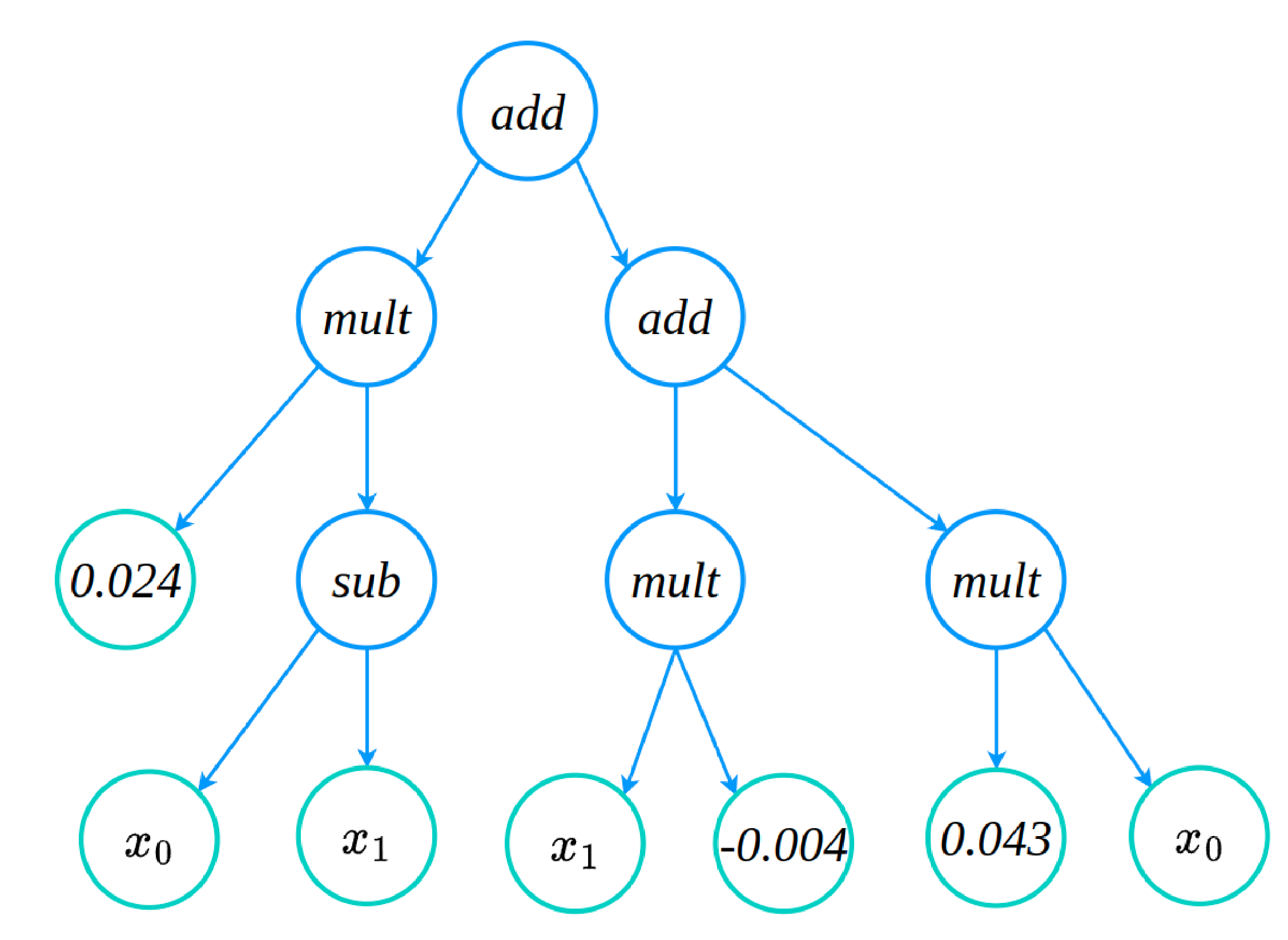}
  \caption{Ultradian Endocrine model: Tree of binary operations recovered for $g_{sym}$.}
  \label{fig:symbolic_tree}
\end{figure}

The gray-box models for $I_p$ and $I_i$ are expressed as
\begin{align}
    \frac{\mathrm{d}I_p}{dt} &= \frac{R_m}{f} + f_{sym}(I_p, I_i)\\
    \frac{\mathrm{d}I_i}{dt} &= g_{sym}(I_p, I_i)
\end{align}
Here, we discover the closed and compact form of $f_{sym}(I_p, I_i)$ and $g_{sym}(I_p, I_i)$ using symbolic regression. In \autoref{table:sbinn}, we present the close and compact form symbolic models for $f_{sym}$ (PINNs and X-TFC) and $g_{sym}$ (PINNs and X-TFC) recovered by using PySR and gplearn. \autoref{table:sbinn} shows a very good agreement between the symbolic models and actual expression represented by semi-discrete system of ODEs. In  \autoref{fig:sr_metric_GI_pysr} and \autoref{fig:sr_metric_GI_gplearn}, we present the plots that show the variation in score and loss against complexity of recovered expression for models learned from X-TFC and PINNs, for PySR and gplearn packages, respectively. Interpretation of the \autoref{fig:sr_metric_GI_pysr} and \autoref{fig:sr_metric_GI_gplearn} are same as those explained in \ref{subsec:SR_PK}. For example, in \autoref{fig:sr_metric_GI_pysr} the convergence with PySR is achieved when the score remains constant while the complexity increases. In \autoref{fig:sr_metric_GI_gplearn}a the convergence with gplearn framework for $f_{sym}$ is achieved at length of expression of 18 and 25 for PINN and X-TFC, respectively. However, for $g_{sym}$ we see that convergence is achieved for length of expression of 13 and 18 for PINN and X-TFC, respectively. In \autoref{fig:symbolic_tree}, we  show the evolved tree of binary operations, obtained from gplearn, in symbolic model recovered for $g_{sym}$ obtained from PINNs. It is to be noted that number of nodes in tree (13) in \autoref{fig:symbolic_tree_pk} represents the the length of expression in the symbolic model. \\

\begin{table}[h!]
\centering
\begin{tabular}{lcc|cc}
\hline
\multicolumn{1}{c}{}  & \multicolumn{2}{c|}{\bf{PySR}}   & \multicolumn{2}{c}{\bf{True expressions}} \\ 
\multicolumn{1}{c}{\textbf{Method}}  & \multicolumn{1}{c}{\bf{$f_{sym}$}}  & \multicolumn{1}{c|}{\bf{$g_{sym}$}}  & \multicolumn{1}{c}{\bf{$f$}} & \multicolumn{1}{c}{\bf{$g$}} 
\\ \hline \hline
X-TFC &   $-0.2333 I_p + 0.0182 I_i$ &  $0.0660 I_p - 0.0280 I_i$  & \multirow{2}{*}{$-0.2333 I_p + 0.0182 I_i$}   &  \multirow{2}{*}{$0.0667 I_p - 0.0282 I_i$}   \\
PINNs &   $-0.2332 I_p + 0.0181 I_i$   & $0.0667 I_p - 0.0282 I_i$ &     &      
\\ \hline \hline
\multicolumn{1}{c}{}  & \multicolumn{2}{c|}{\bf{gplearn}}   & \multicolumn{2}{c}{\bf{True expressions}} \\ 
\multicolumn{1}{c}{\textbf{Method}}  & \multicolumn{1}{c}{\bf{$f_{sym}$}}  & \multicolumn{1}{c|}{\bf{$g_{sym}$}}  & \multicolumn{1}{c}{\bf{$f$}} & \multicolumn{1}{c}{\bf{$g$}} 
\\ \hline \hline
X-TFC &   $-0.2331 I_p + 0.0183 I_i$ &  $0.066 I_p - 0.028 I_i$  &  \multirow{2}{*}{$-0.2333 I_p + 0.0182I_i$}   &  \multirow{2}{*}{$0.0667 I_p - 0.0282 I_i$}   \\
PINNs &   $-0.2329 I_p + 0.0178 I_i$   & $0.068 I_p - 0.029 I_i$ &     &      \\ \hline
\end{tabular}
\caption{Results of symbolic regression for gray-box discovering of Ultradian Endocrine model using the PySR package (top) developed by Cranmer \cite{cranmer2023interpretable} and the method implemented in gplearn \cite{stephens2015gplearn} package (bottom).}
\label{table:sbinn}
\end{table}

\newpage

\section{Summary and Discussion}\label{sec:conclusions}
In this paper, we have presented a comprehensive framework named AI-Aristotle, which combines two neural network-based methods (X-TFC and PINNs) with two symbolic regression techniques to address the challenging tasks of parameter discovery and gray-box identification in Systems Biology problems.

Our framework was evaluated on two benchmark problems: the pharmacokinetics drug absorption model and the ultradian endocrine model describing glucose-insulin interactions. The results demonstrated the capability of both X-TFC and PINNs to accurately estimate parameters even with limited data, showcasing their potential for model calibration in real-world scenarios. In the gray-box identification simulations, our framework successfully discovered the missing terms in the differential equations governing the systems. The learned functions exhibited high accuracy even with a small number of data points. This ability to identify gray-box terms is essential for improving model fidelity and understanding complex systems where some underlying mechanisms are not fully known. We further distilled the learned neural network models using two symbolic regression algorithms, providing interpretable mathematical expressions. This process enhances the transparency and usability of the models, facilitating their integration into scientific research and decision-making processes.

Our study has unveiled a noticeable trend in how dataset size affects the performance of different methods. When we look at the X-TFC method, increasing the number of data points leads to progressively improved results. However, when dealing with relatively small datasets, the PINNs method outperforms on accuracy. This superiority can be attributed to PINNs' efficiency in handling sparse datasets and approximating complex functions with fewer data points. As the dataset size expands, the X-TFC method overtakes PINNs in both accuracy and computational efficiency. In particular, the latter occurs because of the use of least-squares optimization as solver instead of the back-propagation. It seems that the optimization error dominates in PINNs and hence no further improvement can be achieved even for more data points. Thus, when choosing between the X-TFC and PINNs methods, careful consideration of dataset size and required computational time is paramount. \\
We perform the distillation of gray-box models obtained by using PINNs and X-TFC methods. Symbolic regression provided  compact and closed form expression for PINN and X-TFC based surrogates. To show the robustness of recovered symbolic expression, we used PySR and gplearn package, and recovered almost identical expressions for Pharmacokinetics and Ultradian Endocrine model. At the implementation level, we find that PySR is a more robust and efficient framework than gplearn, for example, for the problems we considered here, it takes 10 minutes for PySR on CPU, while it takes up to one hour for gplearn. Also, PySR requires less effort in tuning the hyperparmeters of the model to perform the symbolic regressions. The robustness of PySR is due to the implementation of simulated annealing based mutation of tree of binary expressions \cite{cranmer2023interpretable}, which is not present in the gplearn framework.


\section{Acknowledgements}
This work was supported by the National Institutes of Health (NIH) Spleen grant R01HL154150 and the Office of Naval Research (ONR) Vannevar Bush grant N00014-22-1-2795. All data and codes used in this manuscript are available on GitHub at \href{https://github.com/mariodeflorio/AI-Aristotle}{https://github.com/mariodeflorio/AI-Aristotle}.

\bibliographystyle{ieeetr}
\bibliography{main}

\appendix

\section{Ablation study of X-TFC for Ultradian Endocrine model}

In this ablation study, we evaluate the X-TFC performance for the unknown term discovery test case for the glucose-insulin model, for different distribution of input weights and bias of the neural network. The results reported in Table \ref{tab:xtfc_ablation} are obtained using a dataset of 1800 points, time-steps of length 5 seconds, $N=6$ collocation point per time-step, and 50 neurons, with $tanh$ as activation function. The random distributions and their probability density functions (PDF), in the range $[a,b] = [-1,1]$, selected for this ablation study are the following:

\begin{itemize}
    \item \textbf{Uniform distribution (unifrnd).} PDF: 
    \begin{equation}
        f(x | a, b) = \begin{cases} 
        \frac{1}{b - a} & \text{if } a \leq x \leq b \\
        0 & \text{otherwise}
\end{cases}
    \end{equation}
    
    \item \textbf{Beta distribution (betarnd).} PDF: 
    \begin{equation}
        f(x | \alpha, \beta) = \frac{x^{\alpha-1}(1-x)^{\beta-1}}{B(\alpha, \beta)}
\end{equation}
where \(B(\alpha, \beta)\) is the beta function defined as:
\begin{equation*}
B(\alpha, \beta) = \int_0^1 t^{\alpha-1}(1-t)^{\beta-1} dt
    \end{equation*}
    \item \textbf{Gaussian distribution (normrnd).} PDF: 
    \begin{equation}
        f(x | \mu, \sigma) = \frac{1}{\sigma \sqrt{2\pi}} \exp\left(-\frac{(x - \mu)^2}{2\sigma^2}\right)
    \end{equation}
    \item \textbf{Exponential distribution (exprnd).} PDF:
    \begin{equation}
        f(x | \lambda) = \lambda \exp(-\lambda x)
    \end{equation}
\end{itemize}
From Table \ref{tab:xtfc_ablation}, one can see that the variation in X-TFC performance while varying the distribution of the random weights and bias is minimal. However, we can notice that the uniform distribution is the one that gives the optimal results in terms of error. This is the reason why it has been selected as the random distribution for our simulations in this paper.

\begin{table}[h!]
\centering
\begin{tabular}{lccc|ccc}
\hline
 \multicolumn{7}{c}{\textbf{X-TFC}} \\ \hline
\multicolumn{1}{l}{\textbf{weights and bias}}  & \multicolumn{3}{c|}{\bf{$f(t)$}}   & \multicolumn{3}{c}{\bf{$g(t)$}} \\ 
\multicolumn{1}{l}{\textbf{distribution}}  & \multicolumn{1}{c}{\bf{MAE}}  & \multicolumn{1}{c}{\bf{RMSE}}  & \multicolumn{1}{c|}{\bf{RE}}  & \multicolumn{1}{c}{\bf{MAE}}  & \multicolumn{1}{c}{\bf{RMSE}}  & \multicolumn{1}{c}{\bf{RE}} \\ \hline \hline
unifrnd &  1.34e-04  &  5.60e-04  & 3.38e-05   & 1.37e-03    & 3.95e-03   & 1.54e-03   \\ 
betarnd  & 1.41e-04  & 6.28e-04  &  3.78e-05 & 2.79e-03  &  7.91e-03  &  3.09e-03 \\
normrnd & 1.30e-04  & 5.54e-04  &  3.34e-05 & 3.82e-03  &  1.34e-02 &  5.22e-03    \\
exprnd  &  4.38e-04 &  2.00e-03 & 1.21e-04  & 5.56e-03  &   1.52e-02  &  5.92e-03 \\
\hline
\end{tabular}
\caption{Study on X-TFC performance for the 
 unknown term discovery of glucose-insulin model, for different random distributions in input weights and bias of the neural network. The comparison is made in terms of $MAE$, $RMSE$, and $RE$, for both functions $f(t)$ and $g(t)$.}\label{tab:xtfc_ablation}
\end{table}

Another interesting evaluation on the X-TFC performance for the unknown term discovery test case for the glucose-insulin model, it can be made for different activation functions of the neural network. Since we have ascertained that the uniform random distribution is our optimal distribution, it is the one we use for this study. The results reported in Table \ref{tab:xtfc_ablation_act} are obtained using a dataset of 1800 points, time-steps of length 5 seconds, $N=6$ collocation point per time-step, and 50 neurons. The activation functions used for this test are:
\begin{itemize}
    \item \textbf{Hyperbolic tangent (tanh)}
    \begin{equation}
        \text{tanh}(x) = \frac{e^x - e^{-x}}{e^x + e^{-x}}
    \end{equation}
    \item \textbf{Logistic (sigmoid)}
    \begin{equation}
        \text{sigmoid}(x) = \frac{1}{1 + e^{-x}}
    \end{equation}
    \item \textbf{Sine activation (sine)}
    \begin{equation}
        \text{sine}(x) = \sin(x)
    \end{equation}
    \item \textbf{Inverse tangent (arctan)}
    \begin{equation}
        \text{arctan}(x) = \tan^{-1}(x)
    \end{equation}
    \item \textbf{Softplus}
    \begin{equation}
        \text{softplus}(x) = \ln(1 + e^x)
    \end{equation}
    \item \textbf{Bent Identity}
    \begin{equation}
        \text{bentidentity}(x) = \frac{\sqrt{x^2 + 1} - 1}{2} + x
    \end{equation}
    \item \textbf{Inverse hyperbolic sine (asinh)}
    \begin{equation}
        \text{asinh}(x) = \ln(x + \sqrt{x^2 + 1})
    \end{equation}
    \item \textbf{Softsign}
    \begin{equation}
        \text{softsign}(x) = \frac{x}{1 + |x|}
    \end{equation}
\end{itemize}

Also for this ablation study, from Table \ref{tab:xtfc_ablation_act} we can see that all these activation functions (for the same set of hyperparameteres), allow to very similar performance in terms of error. This proves the robustness of the X-TFC method for both random distributions and activation functions. The choice of selecting $tanh$ as activation function is given by the fact that it returns the best performance for the inversion of the function $g(t)$, which is the most challenging among the two functions to retrieve.

\begin{table}[h!]
\centering
\begin{tabular}{lccc|ccc}
\hline
 \multicolumn{7}{c}{\textbf{X-TFC}} \\ \hline
\multicolumn{1}{l}{\textbf{activation}}  & \multicolumn{3}{c|}{\bf{$f(t)$}}   & \multicolumn{3}{c}{\bf{$g(t)$}} \\ 
\multicolumn{1}{l}{\textbf{function}}  & \multicolumn{1}{c}{\bf{MAE}}  & \multicolumn{1}{c}{\bf{RMSE}}  & \multicolumn{1}{c|}{\bf{RE}}  & \multicolumn{1}{c}{\bf{MAE}}  & \multicolumn{1}{c}{\bf{RMSE}}  & \multicolumn{1}{c}{\bf{RE}} \\ \hline \hline
tanh &  1.34e-04  &  5.60e-04  & 3.38e-05   & 1.37e-03    & 3.95e-03   & 1.54e-03   \\ 
sigmoid  &  3.18e-05 & 1.21e-04  & 7.29e-06
  & 5.20e-03  &  1.76e-02  & 6.85e-03  \\
sine  & 2.50e-05  &  8.63e-05 & 5.20e-06  & 1.91e-02  &  4.67e-02  & 1.82e-02  \\
 arctan & 1.61e-04  &  6.39e-04 & 3.85e-05  &  3.70e-03 &  1.23e-02  &  4.79e-03 \\
 softplus   & 2.54e-05  &  8.77e-05  & 5.29e-06  & 4.81e-03  &   1.64e-02 &  6.41e-03 \\
 bent identity     & 7.74e-05  & 3.72e-04  &  2.24e-05 &  3.84e-03 &  1.30e-02  &  5.09e-03 \\
 asinh  &  1.17e-04  & 4.86e-04  & 2.93e-05  & 3.08e-03  &  9.79e-03  &  3.82e-03 \\
 softsign   &  3.14e-05 & 1.22e-04  & 7.35e-06  & 4.15e-03 &  1.43e-02  & 5.57e-03  \\
\hline
\end{tabular}
\caption{Study on X-TFC performance for the 
 unknown term discovery of glucose-insulin model, for different activation functions in the neural network. The comparison is made in terms of $MAE$, $RMSE$, and $RE$, for both functions $f(t)$ and $g(t)$.}\label{tab:xtfc_ablation_act}
\end{table}

\section{Ablation study of PINNs for Ultradian Endocrine model}

In this comprehensive ablation study of PINNs, we present various scenarios, each characterized by distinct configurations. These configurations differ in terms of the number of neural networks employed, variations in neural network architecture, adjustments to activation functions, and experiments with varying numbers of data points.\\
Our study is categorized into three primary groups, each based on the number of distinct neural networks:

\begin{enumerate}
    \item \textbf{Single Neural Network with Eight Outputs :} In this setup, a single neural network with one input and eight outputs is employed. Six of these outputs represent distinct state variables, while the remaining two outputs represent functions \(f(t)\) and \(g(t)\).
    
    \item \textbf{Two Separated Neural Networks:} In this configuration, two separate neural networks are utilized—one for approximating the six state variables and another for \(f(t)\) and \(g(t)\).
    
    \item \textbf{Three Separated Neural Networks:} This setup involves three distinct neural networks—one for the six state variables, one for \(f(t)\), and one for \(g(t)\).
\end{enumerate}

The outcomes of our experimentation indicated that the first group yielded suboptimal results. The dynamics of the two unknown functions slightly deviated from the primary neural network, leading to limited improvement even with an increased number of iterations. The third group also proved to be ineffective due to extended computational times and slow learning rates, particularly for $g(t)$.

As a result, the second architecture, which employs two separate neural networks for the state variables and $f(t)$ and $g(t)$, demonstrated the most promising results. Consequently, we directed our focused investigation towards this architecture. We made adjustments to the number of collocation points, the number of data points, and the architecture of the second neural network representing $f(t)$ and $g(t)$. Furthermore, we explored different activation functions, including Rowdy (a layer-wise adaptive activation function \cite{jagtap2022deep}), Swish, and Tanh. The performance of PINNs in simulating a 30-hour time span is meticulously assessed and presented in Table \ref{tab:ablattion_PINNs}.\\ 
It's worth mentioning that elimination of either the output scaling layer or the input feature layer causes a decrease in accuracy and can lead to convergence issues, potentially getting stuck in local minima. Hence, we utilized both layers to compare the results.

\begin{table}[h!]
\centering
\begin{tabular}{lccccc}
\hline
\multicolumn{6}{c}{\textbf{PINNs}} \\ \hline
\textbf{Act.} & \textbf{Arch.} & \textbf{\# of Data} & \textbf{\# of iter.} & \textbf{$N_c$}& \textbf{MAE} \\ 
\hline \hline
$Swish$ & 32, 4 & 360, 360 & 1e06 &1800 &1.99e-02, 5.49e-02\\
$Swish$ & 32, 4  & 900, 360 & 1e06 &1800 &1.98e-02, 3.31e-02\\
$Swish$ & 20, 6 & 360, 360 & 1e06 &2400 &3.20e-02, 6.47e-02\\
$Swish$ & 32, 4  & 360, 360 & 1e06 & 2400& 1.87e-02, 5.31e-02 \\
$Tanh$  & 32, 4  & 360, 360 & 1e06 & 2400 &1.06e-01, 1.48e-01\\
$Rowdy$ & 32, 4  & 360, 360 & 1e06 &2400 &1.58e-02, 7.90e-02\\
\hline
\end{tabular}
\caption{Comparison of PINNs with different designs. The first and second numbers in the 'Arch.' column correspond to the width and depth of the second neural network, respectively. The first and second numbers in the '\# of data' column correspond to the number of data points for $G$ and $I_p$, respectively. '$N_c$' corresponds to Number of collocation points. In the 'MAE' column, the values correspond to the computed MAE for $f(t)$ and $g(t)$, respectively.}
\label{tab:ablattion_PINNs}
\end{table}

Performance evaluation is based on the Mean Absolute Error (MAE), calculated as described previously. The PINNs were trained with diverse configurations, encompassing different activation functions, network architectures, the number of data points, and the number of collocation points. The impact of these various configurations on the accuracy of the PINNs can be summarized as follows:

\begin{itemize}
    \item The use of the Swish activation function yielded relatively low MAE, rendering it a favorable choice for precise simulations.
    \item Utilizing an increased number of data points for glucose led to further reductions in MAE for both $f(t)$ and $g(t)$.
    \item However, augmenting the complexity of the architecture (as observed in the 6-layer Swish configuration) did not necessarily result in improved performance, indicating a trade-off between complexity and accuracy.
    \item The Tanh activation function exhibited significantly higher MAE compared to Swish, indicating its limited suitability for this simulation.
    \item The Rowdy activation function demonstrated competitive performance with relatively low MAE, making it a viable option. However, it is essential to note that the Rowdy method entails higher computational costs.
\end{itemize}

These findings underscore the critical role of activation functions and architecture in achieving accurate simulations. The Swish activation function, in conjunction with an appropriate architecture, emerged as the most promising configuration for this simulation.\\
Consequently, we leveraged the outcomes from the second row in Table \ref{tab:ablattion_PINNs} as inputs for the symbolic regression step. The computational time associated with this choice amounted to 4085.87 seconds.


\end{document}